\documentclass[11pt,a4paper]{article}
\usepackage[hyperref]{emnlp-ijcnlp-2019}
\usepackage{times}
\usepackage{latexsym}

\usepackage{tabularx} % extra features for tabular environment
\usepackage{amsmath,amssymb}  % improve math presentation
\usepackage{graphicx} % takes care of graphic including machinery
\usepackage{cite} % takes care of citations
\usepackage{url}
\usepackage{xspace}
\usepackage{algorithmicx,algorithm}
\usepackage{algpseudocode}
\usepackage{mathtools}
\usepackage[inline]{enumitem}
\usepackage{booktabs}
\usepackage{bbm}
\usepackage{colortbl}
\usepackage{stmaryrd}
\usepackage{xcolor}
\usepackage{caption}
\usepackage{subcaption}
\usepackage{arydshln}
\usepackage{pifont}
\usepackage{multirow}
\usepackage{todonotes}
\usepackage{cleveref}
\usepackage{amsthm}
\usepackage{pifont}% http://ctan.org/pkg/pifont
\newcommand{\cmark}{{\color[RGB]{0,128,0}\ding{51}}}%
\newcommand{\xmark}{{\color[RGB]{214,0,0}\ding{55}}}%
\usetikzlibrary{shapes,snakes}
\usetikzlibrary{shapes.geometric}

\aclfinalcopy % Uncomment this line for the final submission

\renewcommand{\vec}[1]{\mathbf{#1}}

\newcommand{\system}{{\fontfamily{pbk}\textsc{Anna}}\xspace}
\newcommand{\problem}{{\fontfamily{pbk}\textsc{Hanna}}\xspace}
\newcommand{\vln}{{\fontfamily{pbk}\textsc{Vln}}\xspace}
\newcommand{\vnla}{{\fontfamily{pbk}\textsc{Vnla}}\xspace}
\newcommand{\cvdn}{{\fontfamily{pbk}\textsc{Cvdn}}\xspace}

\newcommand\myeq{\mathrel{\stackrel{\makebox[0pt]{\mbox{\normalfont\tiny def}}}{=}}}

\newcommand{\aask}{\hat{a}^{\textrm{ask}}}
\newcommand{\anav}{\hat{a}^{\textrm{nav}}}

\newcommand{\testSet}{\textsc{Test}\xspace}
\newcommand{\valSet}{\textsc{Val}\xspace}
\newcommand{\unseenLang}{\textsc{SeenEnv}\xspace}
\newcommand{\unseenAll}{\textsc{UnseenAll}\xspace}

\newcommand{\anavpol}{\hat{\pi}_{\textrm{nav}}}
\newcommand{\aaskpol}{\hat{\pi}_{\textrm{ask}}}
\newcommand{\tnavpol}{\pi_{\textrm{nav}}^\star}
\newcommand{\taskpol}{\pi_{\textrm{ask}}^\star}

\newcommand{\randomWalk}{\textsc{RandomWalk}\xspace}
\newcommand{\forward}{\textsc{Forward10}\xspace}
\newcommand{\shortest}{\textsc{Shortest}\xspace}

\newcommand{\noAsk}{\textsc{NoAsk}\xspace}
\newcommand{\randomAsk}{\textsc{RandomAsk}\xspace}

\DeclarePairedDelimiter\floor{\lfloor}{\rfloor}

\newcommand{\argmin}[1]{\underset{#1}{\textrm{argmin}} \ \ }

\algrenewcommand\alglinenumber[1]{\textcolor{black!50!white}{\tiny #1:}}

\newtheorem{lemma}{Lemma}

\def\equationautorefname~#1\null{Eq~#1\null}
\renewcommand{\sectionautorefname}{\S\kern-0.2em}
\renewcommand{\subsectionautorefname}{\S\kern-0.2em}
\renewcommand{\subsubsectionautorefname}{\S\kern-0.2em}
\makeatletter \newcommand{\ALC@uniqueautorefname}{line} \makeatother

\newcommand{\diamondA}{\tikz\node[diamond,fill={rgb,255:red,255;green,255;blue,0},draw,scale=0.2]{\huge\textsf{A}};\xspace}
\newcommand{\circleB }{\tikz\node[circle ,fill={rgb,255:red,255;green,255;blue,0},draw,scale=0.4]{\textsf{B}};\xspace}
\newcommand{\squareC }{\tikz\node[        fill={rgb,255:red,255;green,255;blue,0},draw,scale=0.5]{\small\textsf{C}};\xspace}
\newcommand{\starEnd }{\tikz\node[star   ,fill={rgb,255:red,255;green,255;blue,0},draw,scale=0.4,star point ratio=2.25]{~};\xspace}
\newcommand{\greenRoute}{\tikz\draw[very thick,color={rgb,255:red,0;green,220;blue,0},scale=0.5](0,0) .. controls (0,0.5) and (0.5,0) .. (0.5,0.5);\xspace}
\newcommand{\cyanRoute}{\tikz\draw[very thick,color={rgb,255:red,0;green,220;blue,220},scale=0.5](0,0) .. controls (0,0.5) and (0.5,0) .. (0.5,0.5);\xspace}

\title{Help, Anna! Visual Navigation with Natural Multimodal Assistance via Retrospective Curiosity-Encouraging Imitation Learning}

\newcommand{\idcs}{${}^\odot$}

\newcommand{\idmsr}{${}^\heartsuit$}

\definecolor{myred}{HTML}{ff1458}

\author{Khanh Nguyen\idcs \and Hal Daum{\'e} III\idcs\idmsr \\
  University of Maryland, College Park\idcs, Microsoft Research, New York\idmsr\\
  \tt{ kxnguyen@cs.umd.edu \ \ hal@umiacs.umd.edu }
 }

\date{}

\begin{document}

\maketitle

\noindent
{\color{myred} This paper is the arXiv version of the paper that appears in the proceedings of EMNLP 2019. The content of the main paper is the exactly same as in the proceedings (modulo citation updates). However, the evaluation method used to obtain the results in the main paper unfortunately induces non-deterministic agent behavior, which makes comparisons difficult. We provide additional results herein obtained via a deterministic evaluation scheme in \autoref{sec:batch_reset}. All conclusions and qualitative claims made in the main paper are unaffected by this change of evaluation scheme, and still hold on the new results. \textbf{We strongly recommend future work reference results in \autoref{sec:batch_reset} when comparing with our methods.}}

\vspace{0.3cm}

A video demo of \problem is available at: 
\vspace{-0.1cm}
\begin{center}
\color{myred}\url{https://youtu.be/18P94aaaLKg}
\end{center}

\vspace{0.6cm}

\begin{abstract}
    %Mobile agents that are capable of leveraging help from humans can accomplish complex navigation tasks in completely unacquainted environments.
%Teaching agents to intelligently request and interpret human assistance requires first developing realistic research platforms. 
%We present \textit{Help, Anna!} (\problem), an interactive vision-based navigation problem in which a mobile agent fulfills object-finding tasks in simulated photorealistic environments by requesting and interpreting natural multimodal assistance. 
%To simulate how humans typically provides assistance, we introduce the concept of Automatic Natural Navigation Assistant (\system), an automated system that aids the agent by request at certain locations in the environment via natural language and visual instructions. We present practical approaches to tackling the \problem problem, including a hierarchical recurrent architecture that models multiple levels of decision-making, and a novel imitation learning algorithm that teaches the agent to avoid repeating past mistakes and predict chance of making progress in the future. 
%Empirical results show that our approaches are more effective than competitive baselines on both seen and unseen environments.

Mobile agents that can leverage help from humans can potentially accomplish more complex tasks than they could entirely on their own.
We develop ``Help, Anna!" (\problem), an interactive photo-realistic simulator in which an agent fulfills object-finding tasks by requesting and interpreting natural language-and-vision assistance. 
An agent solving tasks in a \problem environment can leverage simulated human assistants, called \system (Automatic Natural Navigation Assistants), which, upon request, provide natural language and visual instructions to direct the agent towards the goals. 
To address the \problem problem, we develop a memory-augmented neural agent that hierarchically models multiple levels of decision-making, and an imitation learning algorithm that teaches the agent to avoid repeating past mistakes while simultaneously predicting its own chances of making future progress. 
Empirically, our approach is able to ask for help more effectively than competitive baselines and, thus, attains higher task success rate on both previously seen and previously unseen environments. We publicly release code and data at {\url{https://github.com/khanhptnk/hanna}} .

%----------------------------------

%Proposed rewrite:

%Mobile agents that can leverage help from humans can potentially accomplish more complex tasks than they could entirely on their own.
%We develop \textit{Help, Anna!} (\problem), a simulated photo-realistic interactive environment in which an agent fulfills object-finding tasks by requesting and interpreting language and vision assistance. 
%An agent solving the \problem problem can leverage a simulated assistant (the Automatic Natural Navigation Assistant, \system), that, upon request, provides this multimodal assistance.
%To address the \problem problem, we develop a hierarchical recurrent neural agent that models multiple levels of decision-making, and a novel imitation learning algorithm that teaches the agent to avoid repeating past mistakes and predict its own chances of making progress in the future. 
%Empirical results show that our approaches---which leverage the ability to ask for help---are more effective than competitive baselines on both previously seen and previously unseen environments.

\end{abstract}

\section{Introduction} \label{sec:intro}
\begin{figure*}[t!]
    \centering
    \includegraphics[width=0.95\textwidth]{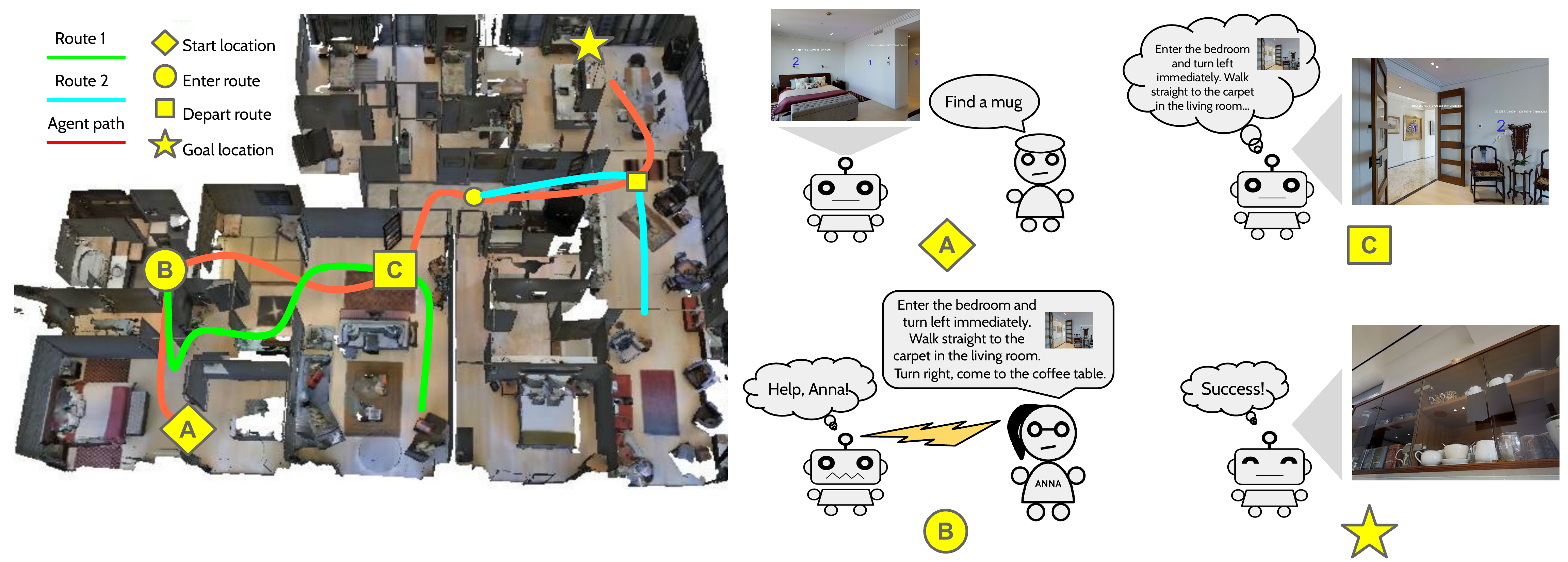}\vspace{-1em}
    \caption{\label{fig:example}An example \problem task. 
    Initially, the agent stands in the bedroom at \diamondA and is requested by a human requester to ``find a mug.'' The agent begins, but gets lost somewhere in the bathroom. It gets to the start location of route \greenRoute (\circleB) to request help from \system. Upon request, \system assigns the agent a navigation subtask described by a natural language instruction that guides the agent to a target location, and an image of the view at that location. The agents follows the language instruction and arrives at \squareC, where it observes a match between the target image and the current view, thus decides to depart route \greenRoute. After that, it resumes the main task of finding a mug. From this point, the agent gets lost one more time and has to query \system for another subtask that helps it follow route \cyanRoute and enter the kitchen. The agent successfully fulfills the task it finally stops within $\epsilon$ meters of an instance of the requested object (\starEnd).
    Here, the \system feedback is simulated using two pre-collected language-assisted routes (\greenRoute and \cyanRoute).
    } \vspace{-0.5em}
\end{figure*}

%Agents that understand assistance in forms of verbal and/or visual instructions can fulfill complex tasks such as cooking (CITE Yang and Yannis), playing video games (CITE), captioning images (CITE), chatting (CITE).
The richness and generalizability of natural language makes it an effective medium for directing mobile agents in navigation tasks, even in environments they have never encountered before \citep{anderson2018vision,chen2018touchdown,misra2018goal,de2018talk,qi2019rerere}. 
Nevertheless, even with language-based instructions, such tasks can be overly difficult for agents on their own, especially in unknown environments.
To accomplish tasks that surpass their knowledge and skill levels, agents must be able to actively seek for and leverage assistance in the environment.
% develop research platform forDrawiningg inspiration from how humans asks for help from others when they cannot accomplish a task by themselves, we empower navigation agents with the capabilities of actively requesting and interpreting human assistance. 
%We posit that \emph{agents that can leverage assistance from humans will be able to accomplish tasks that surpass their knowledge and skill levels}.
Humans are rich external knowledge sources but, unfortunately, they may not be available all the time to provide guidance, or may be unwilling to help too frequently.
%They can even be upset and distrust an agent if it becomes too reliant on them. 
To reduce the needed effort from human assistants, it is essential to design research platforms for teaching agents to request help mindfully.
%We construct a research platform with simulated human assistance to study this problem.

%\hal{I've gone through and tried to remove as many widows/orphans as i can find, but please as you edit things make sure not to introduce no ones, and if i missed any, please fix them.}
%\kcomment{You mean, for example, the phrase "he or she"?}

In natural settings, human assistance is often:
\begin{itemize}[label={$\diamond$},nolistsep]
\item derived from \textit{interpersonal} interaction (a lost tourist asks a local for directions);
\item \textit{reactive} to the situation of the receiver, based on the assistant's knowledge (the local may guide the tourist to the goal, or may redirect them to a different source of assistance);
\item delivered via a \textit{multimodal} communication channel (the local uses a combination of language, images, maps, gestures, etc.).
\end{itemize}
\noindent
We introduce the ``Help, Anna!'' (\problem) problem (\autoref{sec:problem}), in which a mobile agent has to navigate (without a map) to an object by interpreting its first-person visual perception and requesting help from \textit{Automatic Natural Navigation Assistants} (\system).
\problem models a setting in which a human is not always available to help, but rather that human assistants are scattered throughout the environment and provide help upon request (modeling the \emph{interpersonal} aspect).
The assistants are \emph{not} omniscient: they are only familiar with certain regions of the environment and, upon request, provide \emph{subtasks}, expressed in language and images (modeling the \emph{multimodal} aspect), for getting \emph{closer} to the goal, not necessarily for fully completing the task (modeling the \emph{reactive} aspect).
%If the agent can directly solve this navigation task, it never needs to request assistance from \system

In \problem, when the agent gets lost and becomes unable to make progress, it has the option of requesting assistance from \system.
%Toward the goal of developing agents that can leverage such assistance, we first need to build research testbeds for fast prototyping and evaluating new methods. 
%The goal of this work is to teach agents to intelligently leverage assistance that emulates those aspects of human assistance. 
At test time, the agent must decide where to go and whether to request help from \system without additional supervision.
At training time, we leverage imitation learning to learn an effective agent, both in terms of navigation, and in terms of being able to decide when it is most worthwhile to request assistance.

This paper has two primary contributions:
\begin{enumerate}[nolistsep,noitemsep]
\item Constructing the \problem simulator by augmenting an indoor photo-realistic simulator with simulated human assistance, mimicking a scenario where a mobile agent finds objects by asking for directions along the way (\autoref{sec:problem}).
\item An effective model and training algorithm for the \problem problem, which includes a hierarchical memory-augmented recurrent architecture that models human assistance as \emph{sub-goals} (\autoref{sec:architecture}), and introduces an imitation learning objective that enhances exploration of the environment and interpretability of the agent's help-request decisions. (\autoref{sec:imitation}).
\end{enumerate}
\noindent
We embed the \problem problem in the photo-realistic Matterport3D environments \citep{chang2017matterport3d} with \emph{no} extra annotation cost by reusing the pre-existing Room-to-Room dataset \citep{anderson2018vision}.
Empirical results (\autoref{sec:results}) show that our agent can effectively learn to request and interpret language and vision instructions, given a training set of 51 environments and less than 9,000 language instructions. 
Even in new environments, where the scenes and the language instructions are previously unseen, the agent successfully accomplishes 47\% of its tasks.
Our methods for training the navigation and help-request policies outperform competitive baselines by large margins.

\section{Related work} \label{sec:related}
Simulated environments provide an inexpensive platform for fast prototyping and evaluating new ideas before deploying them into the real world. 
Video-game and physics simulators are standard benchmarks in reinforcement learning \citep{todorov2012mujoco,mnih2013playing,kempka2016vizdoom,brockman2016openai,vinyals2017starcraft}.
Nevertheless, these environments under-represent the complexity of the world.
Realistic simulators play an important role in sim-to-real approaches, in which an agent is trained with arbitrarily many samples provided by the simulators, then transferred to real settings using sample-efficient transfer learning techniques \citep{kalashnikov2018qt,andrychowicz2018learning,karttunen2019video}. 
While modern techniques are capable of simulating images that can convince human perception \citep{karras2017progressive,karras2018style}, simulating language interaction remains challenging. 
There are efforts in building complex interactive text-based worlds \citep{cote18textworld,urbanek2019learning} but the lack of a graphical component makes them not suitable for visually grounded learning.
On the other hand, experimentation on real humans and robots, despite expensive and time-consuming, are important for understanding the true complexity of real-world scenarios \citep{chai2018language, chai2016collaborative,rybski2007interactive,mohan2014learning,liu2016jointly,she2014back}.

Recent navigation tasks in photo-realistic simulators have accelerated research on teaching agents to execute human instructions. Nevertheless, modeling human assistance in these problems remains simplistic (\autoref{tab:compare_problem}): they either do not incorporate the ability to request additional help while executing tasks \citep{misra2014tell,misra2017mapping,anderson2018vision,chen2018touchdown,das2018embodied,misra2018goal,wijmans2019embodied,qi2019rerere}, or mimic human verbal assistance with primitive, highly scripted language \citep{nguyen2019vision,chevalier2018babyai}. 
\problem improves the realisticity of the \vnla setup \citep{nguyen2019vision} by using fully natural language instructions. 
\citet{Tellex2014AskingFH} introduce inverse semantics for teaching robots to request intervention from humans but, in their setup, the robot specifies subtasks for the humans to execute, whereas our setup is opposite: the humans specify subtasks for the robot to execute.

Imitation learning algorithms are a great fit for training agents in simulated environments: access to ground-truth information about the environments allows optimal actions to be computed in many situations.
The ``teacher'' in standard imitation learning algorithms \citep{daume2009search,ross2011reduction,ross2014reinforcement,chang2015learning,sun2017deeply,sharaf2017structured} does not take into consideration the agent's capability and behavior.
\citet{he2012imitation} present a coaching method where the teacher gradually increases the complexity of its demonstrations over time. 
\citet{welleck2019neural} propose an "unlikelihood" objective, which, similar to our curiosity-encouraging objective, penalizes likelihoods of candidate negative actions to avoid mistake repetition.
Our approach takes into account the agent's past and future behavior to determine actions that are most and least beneficial to them, combining the advantages of both model-based and progress-estimating methods \citep{wang2018look,ma2019self,ma2019regretful}.

\section{The \problem Simulator} \label{sec:problem}
\begin{table}[t!]
    \small
    \centering
    \setlength{\tabcolsep}{1.6pt}
	\begin{tabular}{l@{~~~}c@{~~~}c@{~~~}c}
		\toprule[1pt]
		  & Request  & Multimodal & Simulated \\
		 Problem & assistance & instructions & humans \\
		 \cmidrule(lr){1-1} \cmidrule(lr){2-2} \cmidrule(lr){3-3} \cmidrule(lr){4-4}
		 \vln & \xmark & \xmark & \xmark \\
		 \vnla  & \cmark & \xmark & \cmark \\
		 \cvdn  & \cmark & \xmark & \xmark \\
		 \problem (this work)  & \cmark & \cmark & \cmark \\
        \bottomrule[1pt] 
	\end{tabular}
	\caption{Comparing \problem with other photo-realistic navigation problems. \vln \citep{anderson2018vision} does not allow agent to request help. \vnla \citep{nguyen2019vision} models an advisor who is always present to help but speaks simple, templated language. \cvdn \citep{thomason2019vision} contains natural conversations in which a human assistant aids another human in navigation tasks but offers limited language interaction simulation, as language assistance is not available when the agent deviates from the collected trajectories and tasks. \problem simulates human assistants that provide language-and-vision instructions that adapt to the agent's current position and goal.}
	\label{tab:compare_problem}
\end{table}

\paragraph{Problem.} \problem simulates a scenario where a \textit{human requester} asks a mobile \textit{agent} via language to find an object in an indoor environment.
The task request is only a high-level command (``find [object(s)]"), modeling the general case when the requester does not need know how to accomplish a task when requesting it.
We assume the task is always feasible: there is at least an instance of the requested object in the environment. 

\autoref{fig:example}, to which references in this section will be made, illustrates an example where the agent is asked to ``find a mug.''
The agent starts at a random location (\diamondA), is given a task request, and is allotted a budget of $T$ time steps to complete the task. 
%The \textit{delegate location} of an object instance is defined as the location of the closest graph node that is in the same room with it. 
The agent succeeds the if its final location is within $\epsilon_{\textrm{success}}$ meters of the location of any instance of the requested object (\starEnd). 
The agent is not given any sensors that help determine its location or the object's location and must navigate only with a monocular camera that captures its first-person view as an RGB image (e.g., image in the upper right of \autoref{fig:example}). 

%\hal{i would lead with a very concrete example, then fill in the details around that example. maybe use the "find a knife" guest example. probably have a figure. say that the agent receives this instruction, does not know what to do, and then goes and finds a human assistant to ask. it says "help me!" and then the human assistant says (and use a real example here from the domain) something like "find your way to the kitchen" and shows a picture of the kitchen (again, show the actual image). then this introduces a subtask that the agent will try to achieve and, when it does, can continue to get a knife (if it knows how) or find another assistant to ask for help again. then after this example, go into the details.}
%\kcomment{you are right!}

The only source of help the agent can leverage in the environment is \textit{assistants}, who are present at both training and evaluation time.
The assistants are not aware of the agent unless it enters their \textit{zones of attention}, which include all locations within $\epsilon_{\textrm{attn}}$ meters of their locations.  %\hal{is this the same $\epsilon$ as before?}
%\kcomment{we can add subscripts, in practice, i use the same values for both} \hal{eh... if they're the same, it's fine to leave as is i think}
When the agent is in one of these zones, it has an option to request help from the corresponding assistant. 
The assistant helps the agent by giving a \textit{subtask}, described by a natural language instruction that guides the agent to a specific location, and an image of the view at that location.

In our example, at \circleB, the assistant says \textsl{\small``Enter the bedroom and turn left immediately. Walk straight to the carpet in the living room. Turn right, come to the coffee table.''} and provides an image of the destination in the living room.
Executing the subtask may not fulfill the main task, but is guaranteed to get the agent to a location closer to a goal than where it was before (e.g., \squareC). 
%\kcomment{there can be multiple goal locations}
%The agent can request help multiple times whenever help is available, even when it is already being assisted.
%The location of the assistants and the object instances are static. 

\paragraph{Photo-realistic Navigation Simulator.}
\problem uses the Matterport3D simulator \citep{chang2017matterport3d,anderson2018vision} to photo-realistically emulate a first-person view while navigating in indoor environments.
\problem features 68 Matterport3D environments, each of which is a residential building consisting of multiple rooms and floors. 
%The agent is equipped with a monocular camera that captures its first-person view as an RGB image. 
Navigation is modeled as traversing an undirected graph $G = (V,E)$, where each location corresponds to a node $v \in V$ with 3D-coordinates $\vec x_v$, and edges are weighted by their lengths (in meters).
%Two locations are navigable from one to the other if they are connected by a path on the graph.  
%The distance between them is the length of the weighted shortest path connecting them, where the weight of each edge is the Euclidean distance between the two locations incident to it. 
%Unless otherwise specified, whenever we refer to a ``distance'' between two locations, we mean shortest-path distances in this graph. 
%For example, ``A is within X meters of B'' means that ``the length of the shortest path between A and B are less than or equal to X meters". 
The state of the agent is fully determined by its pose $\tau = (v, \psi, \omega)$, where $v$ is its location, $\psi \in (0, 2\pi]$ is its heading (horizontal camera angle), and $\omega \in \left[-\frac{\pi}{6}, \frac{\pi}{6}\right]$ is its elevation (vertical camera angle). The agent does not know $v$, and the angles are constrained to multiples of $\frac \pi 6$. %\hal{it's not clear to me if the agent knows $\tau$ or if only the simulator does.}
%\kcomment{it is a bit complex. the agent does not know $v$ but it can keep track of the camera angles. in practice, the camera angles is not an input} \hal{in that case i'd just say "of which the agent does not know $v$" or sth like that}
%The camera can be rotated horizontally or vertically by a multiple of 30 degrees. 
In each step, the agent can either stay at its current location, or it can rotate toward and go to a location adjacent to it in the graph\footnote{We use the ``panoramic action space'' \citep{fried2018speaker}.}.
%If the agent decides to move, it first rotates the camera so that the center of its view is closest to the coordinates of the next location. 
%Given a pose, the simulator computes an RGB image representing the corresponding first-person view. 
Every time the agent moves (and thus changes pose), the simulator recalculates the image to reflect the new view. % \hal{i'm not 100\% clear on exactly what the action space is. can you be more pedantic?}
%\kcomment{we will specify it when describe the agent policies. do you want to do it here?} \hal{can you point me to where it's defined? i don't see it.}
%\kcomment{it is in the beginning of section 4}

%The agent tackles this problem by learning human vision-based navigation and grounded language-understanding capabilities.
%First, it has to ground the assisting instructions from its visual perception, and translate them into sequences of actions.
%Second, it must be able recognize situations where additional assistance is required to make progress. 
%As the agent relies entirely on its own intelligence and help from humans, this human-inspired approach requires no special infrastructure to be installed in the environment.

\paragraph{Automatic Natural Navigation Assistants (\system).} 
\system is a simulation of human assistants who do not necessarily know themselves how to optimally accomplish the agent's goal: they are only familiar with scenes along certain paths in the environment, and thus give advice to help the agent make partial progress.
Specifically, the assistance from \system is modeled by a set of \emph{language-assisted routes} $R = \{ r_1, r_2, \ldots, r_{|R|} \}$.
Each route $r = (\psi^r, \omega^r, p^r, l^r)$ is defined by initial camera angles $(\psi^r, \omega^r)$, a path $p^r$ in the environment graph, and a natural language instruction $l^r$.
A route becomes \textit{enterable} when its start location is adjacent to and within $\epsilon_{\textrm{attn}}$ meters of the agent's location.
When the agent enters a route, it first adjusts its camera angles to $(\psi^r, \omega^r)$, then attempts to interpret the language instructions $l^r$ to traverse along $p^r$. 
At any time, the agent can \emph{depart} the route by stopping following $l^r$. 
An example of a route in \autoref{fig:example} is the combination of the initial camera angles at \circleB, the path \greenRoute, and the language instruction \textsl{\small``Enter the bedroom and turn left immediately\dots''}

The set of all routes starting from a location simulates a \emph{human assistant} who can recall scenes along these routes' paths. 
The \emph{zone of attention} of the simulated human is the set of all locations from which the agent can enter one of the routes;
when the agent is in this zone, it may ask the human for help. 
Upon receiving a help request, the human selects a route $r^{\star}$ for the agent to enter (e.g., \greenRoute), and a location $v_d$ on the route where it wants the agent to depart (e.g., \squareC).
It then replies the agent with a \emph{multimedia message} $(l^{r^{\star}}, \mathcal{I}^{v_d})$, where $l^{r^{\star}}$ is the selected route's language instruction, and $\mathcal{I}^{v_d}$ is an image of the panoramic view at the departure location.
The message describes a \emph{subtask} which requires the agent to follow the direction described by $l^{r^{\star}}$ and to stop if it reaches the location referenced by $\mathcal{I}^{v_d}$.
The route $r^{\star}$ and the departure node $v_d$ are selected to get the agent as close to a goal location as possible. 
Concretely, let $R_{\textrm{curr}}$ be the set of all routes associated with the requested human.
The selected route minimizes the distance to the goal locations among all routes in $R_{\textrm{curr}}$:
\begin{align}
    r^{\star} &= \underset{r \in R_{\textrm{curr}}}{\mathrm{argmin}} \ \ \bar{d}\left(r, V_{\textrm{goal}}\right)\\
    \textrm{where } & \bar{d}\left( r, V_{\textrm{goal}} \right) \myeq \min_{g \in V_{\textrm{goal}}, v \in p^r } d \left( g, v \right)
    %r^{\star} &= \underset{r \in R_{\textrm{curr}}}{\mathrm{argmin}} \min_{g \in V_{\textrm{goal}}, v \in p^r } d \left( g, v \right)
\end{align} $d(.,.)$ returns the (shortest-path) distance between two locations, and $V_{\textrm{goal}}$ is the set of all goal locations.
The departure location minimizes the distance to the goal locations among all locations on the selected route:
\begin{align}
    v_d 
    &= \underset{g \in V_{\textrm{goal}}, v \in p_{r^{\star}}}{\mathrm{argmin}} \ \  d\left( g, v \right) \myeq \bar{d}\left(r^{\star}, V_{\textrm{goal}} \right)
\end{align}
When the agent chooses to depart the route (not necessarily at the departure node), the human further assists it by providing $\mathcal{I}^{g^{\star}}$, an image of the panoramic view at the goal location closest to the departure node:
\begin{align}
     g^{\star} = \argmin{g \in V_{\textrm{goal}}} d\left(g, v_d \right)
\end{align} 

The way the agent leverages \system to accomplish tasks is analogous to how humans travel using public transportation systems (e.g., bus, subway). 
For example, passengers of a subway system utilize fractions of pre-constructed routes to make progress toward a destination. 
They execute travel plans consisting of multiple subtasks, each of which requires entering a start stop, following a route (typically described by its name and last stop), and exiting at a departure stop (e.g., \textsl{\small``Enter the Penn Station, hop on the Red line in the direction toward the South Ferry, get off at the World Trade Center"}). 
Occasionally, users walk short distances (at a lower speed) to switch routes. 
Our setup follows the same principle, but instead of having physical vehicles and railways, we employ low-level language-and-vision instructions as the ``high-speed means" to accelerate travel.

\paragraph{Constructing \system route system.} Given a photo-realistic simulator, the primary cost for constructing the \problem problem comes from crowd-sourcing the natural language instructions. 
Ideally, we want to collect sufficient instructions to simulate humans in any location in the environment. 
Let $N = |V|$ be the number of locations in the environment. 
Since each simulated human is familiar with at most $N$ locations, in the worst case, we need to collect $O(N^2)$ instructions to connect all location pairs. 
However, we theoretically prove that, assuming the agent executes instructions perfectly, it is possible to guide the agent between any location pair by collecting only $\Theta(N \log N)$ instructions.
The key idea is using $O(\log N)$ instead of a single instruction to connect each pair, and reusing an instruction for multiple routes. 

\begin{lemma} (proof in Appendix A) To guide the agent between any two locations using $O(\log N)$ instructions, we need to collect instructions for $\Theta(N \log N)$ location pairs.
\end{lemma}

In our experiments, we leverage the pre-existing Room-to-room dataset \citep{anderson2018vision} to construct the route system. 
This dataset contains 21,567 natural language instructions crowd-sourced from humans and is originally intended to be used for the Vision-Language Navigation task (such as those in \autoref{fig:example}), where an agent executes a language instruction to go to a location.
We exclude instructions of the test split and their corresponding environments because ground-truth paths are not given. 
We use (on average) 211 routes to connect (on overage) 125 locations per environment. 
Even though the routes are selected randomly in the original dataset, our experiments show that they are sufficient for completing the tasks (assuming perfect assistance interpretation). 

%On overage, there are X locations and Y instructions per environment, so the number of instructions almost is linear to ... 
%It is worth noting that similar datasets have been developed for several simulated environments such as CITE. 
%The \problem problem can potentially be embedded in these environments with minimal extra annotation cost. 
%\hal{don't forget to fill this in}

\section{Retrospective Curiosity-Encouraging Imitation Learning}  \label{sec:imitation}
\newcommand{\algopar}{\par\hskip\algorithmicindent}

\begin{algorithm}[t!]
\caption{Task episode, given agent help-request policy $\aaskpol$ and navigation policy $\anavpol$}
\label{alg:task_ep}
\begin{algorithmic}[1]
\small
\State agent receives task request $e$
\State initialize the agent mode: $m \gets \texttt{main\_task}$
\State initialize the language instruction: $l_0 \gets e$
\State initialize the target image: $I^{\textrm{tgt}}_0 \gets \texttt{None}$
\For {$t = 1 \dots T$}
    \State let $s_t$ be the current state, $o_t$ the current observation, \algopar and $\tau_t = (v_t, \psi_t, \omega_t)$ the current pose %, which includes the current pose $\tau_t = (v_t, \psi_t, \omega_t)$.
%    \State $o_t$ is the current observation. 
    \State agent makes a help-request decision $\aask_t
    \sim \aaskpol(o_t)$
    \State carry on task from the previous step: \algopar $l_t \gets l_{t - 1}, I^{\textrm{tgt}}_t = I^{\textrm{tgt}}_{t - 1}$
    \If {$\aask_t = \texttt{request\_help}$}
        \State set mode: $m \gets \texttt{sub\_task}$
        \State request help: $(r, I^{\textrm{depart}}, I^{\textrm{goal}}) \gets \system(s_t)$
        \State set the language instruction: $l_t \gets l^r$
        \State set the target image: $I^{\textrm{tgt}}_t \gets I^{\textrm{depart}}$
        \State set the navigation action: \algopar $\anav_t \gets (p^r_0, \psi^r - \psi_t, \omega^r - \omega_t)$, \algopar where $p_0^r$ is the start location of route $r$
        %\hal{the variables here aren't in scope. where do they come from?}
        %\kcomment{they were there before. let me add them back}
    \Else 
        \State agent chooses navigation: $\anav_t \sim \anavpol(o_t)$
        \If {$\anav_t = \texttt{stop}$}
        \If {$m = \texttt{main\_task}$}
            \State \textbf{break}
        \Else
            \State set mode: $m \gets \texttt{main\_task}$
            \State set the language instruction: $l_t \gets e$
            \State set the target image: $I^{\textrm{tgt}}_t \gets I^{\textrm{goal}}$
             \State set navigation action: $\anav_t \gets (v_t, 0, 0) $
        \EndIf
    \EndIf
    \EndIf
    \State agent executes $\anav_t$ to go to the next location %\hal{what if $\anav_t$ is stop but we hit the else branch above?}
    %\kcomment{solved}
\EndFor
\end{algorithmic}
\end{algorithm}

\noindent\textbf{Agent Policies}. Let $s$ be a fully-observed state that contains ground-truth information about the environment and the agent (e.g., object locations, environment graph, agent parameters, etc.). 
Let $o_s$ be the corresponding observation given to the agent, which only encodes the current view, the current task, and extra information that the agent keeps track of (e.g., time, action history, etc.).
%\hal{why not call this $o$ for observation in keeping with POMDP literature?}
%\kcomment{done}
The agent maintains two stochastic policies: a navigation policy $\anavpol$ and a help-request policy $\aaskpol$. 
Each policy maps an observation to a probability distribution over its action space. 
Navigation actions are tuples $(v, \Delta \psi, \Delta \omega)$, where $v$ is a next location that is adjacent to the current location and $(\Delta \psi, \Delta \omega)$ is the camera angle change. %(the latter of which is deterministic given the former).
A special \texttt{stop} action is added to the set of navigation actions to signal that the agent wants to terminate the main task or a subtask (by departing a route).
The action space of the help-request policy contains two actions: \texttt{request\_help} and \texttt{do\_nothing}.
The \texttt{request\_help} action is only available when the agent is in a zone of attention.
\autoref{alg:task_ep} describes the effects of these actions during a task episode.

\paragraph{Imitation Learning Objective.}
The agent is trained with imitation learning to mimic behaviors suggested by a navigation teacher $\tnavpol$ and a help-request teacher $\taskpol$, who have access to the fully-observed states.
In general, imitation learning \citep{daume2009search,ross2011reduction,ross2014reinforcement,chang2015learning,sun2017deeply} finds a policy $\hat{\pi}$ that minimizes the expected imitation loss $\mathcal{L}$ with respect to a teacher policy $\pi^\star$ under the agent-induced state distribution $\mathcal{D}_{\hat{\pi}}$:
\begin{align}
    \min_{\hat{\pi}} \mathbb{E}_{s \sim \mathcal{D}_{\hat{\pi}}}\left[ \mathcal{L}(s, \hat{\pi}, \pi^\star) \right]
\end{align}
We frame the \problem problem as an instance of \emph{Imitation Learning with Indirect Intervention} (I3L) \citep{nguyen2019vision}.
Under this framework, assistance is viewed as augmenting the current environment with new  information. Interpreting the assistance is cast as finding the optimal acting policy in the augmented environment. 
Formally, I3L searches for policies that optimize:
\begin{align}
    \min_{\aaskpol, \anavpol} & \mathbb{E}_{s \sim \mathcal{D}^{\textrm{state}}_{\anavpol, \mathcal{E}}, \mathcal{E} \sim  \mathcal{D}^{\textrm{env}}_{\aaskpol}}\left[ L(s) \right] \\
    L(s) & =  \mathcal{L}_{\textrm{nav}}(s, \anavpol, \tnavpol) + \mathcal{L}_{\textrm{ask}}(s, \aaskpol, \taskpol) \nonumber
\end{align} where $\mathcal{L}_{\textrm{nav}}$ and $\mathcal{L}_{\textrm{ask}}$ are the navigation and help-request loss functions, respectively, $\mathcal{D}^{\textrm{env}}_{\aaskpol}$ is the environment distribution induced by $\aaskpol$, and $\mathcal{D}^{\textrm{state}}_{\anavpol, \mathcal{E}}$ is the state distribution induced by $\anavpol$ in environment $\mathcal{E}$.
A common choice for the loss functions is the agent-estimated negative log likelihood of the reference action:
\begin{align}
    \mathcal{L}_{\textrm{NL}}(s,\hat{\pi}, \pi^{\star}) = -\log \hat{\pi}(a^\star \mid o_s)
\label{eqn:nl_loss}
\end{align} where $a^\star$ is the reference action suggested by $\pi^\star$. 
We introduce novel loss functions that enforce more complex behaviors than simply mimicking reference actions. 

%\kcomment{don't want to say $\pi(s)$ because the teacher can return multiple things}
%\hal{should this be "the reference actionS suggested" then? but then what does that mean? do you sum over?}
%\kcomment{we draw $s$ from the agent-induced state distribution $\mathcal{D}_{\pi}$ so no need to sum. the expectation handles the sum.}

\paragraph{Reference Actions.} The navigation teacher suggests a reference action $a^{\textrm{nav}\star}$ that takes the agent to the next location on the shortest path from its location to the target location.
Here, the target location refers to the nearest goal location (if no target image is available), or the location referenced by the target image (provided by \system). 
If the agent is already at the target location, $a^{\textrm{nav}\star} = \texttt{stop}$.
To decide whether the agent should request help, the help-request teacher verifies the following conditions:
\begin{enumerate}[nolistsep]
    \item \texttt{lost}: the agent will not get (strictly) closer to the target location in the future; 
    \item \texttt{uncertain\_wong}: the entropy\footnote{Precisely, we use \emph{efficiency}, or entropy of base $|A^{\textrm{nav}}| = 37$, where $A^{\textrm{nav}}$ is the navigation action space. } of the navigation action distribution is greater than or equal to a threshold $\gamma$, and the highest-probability predicted navigation action is \emph{not} suggested by the navigation teacher;
    \item \texttt{never\_asked}: the agent previously never requested help at the current location; 
\end{enumerate}
If condition (1) or (2), and condition (3) are satisfied, we set $a^{\textrm{ask}\star} = \texttt{request\_help}$; otherwise,  $a^{\textrm{ask}\star} = \texttt{do\_nothing}$.

%We augment this loss function with new auxiliary losses that enforce more complex behaviors.
%Hence, in addition to reference actions, the teachers provide extra information needed for computing the auxiliary losses.

\paragraph{Curiosity-Encouraging Navigation Teacher.}
In addition to a reference action, the navigation teacher returns $A^{\textrm{nav}\otimes}$, the set of all non-reference
actions that the agent took at the current location while executing the same language instruction:
\begin{align}
    A^{\textrm{nav}\otimes}_t = &\big\{ a \in A^{\textrm{nav}} :  \exists t' < t, ~~v_t = v_{t'}, \\&\quad\quad~l_t = l_{t'}, ~~a = a_{t'}^{\textrm{nav}} \neq a^{\textrm{nav}\star}_{t'} \big\} \nonumber
\end{align} where $A^{\textrm{nav}}$ is the navigation action space.

We devise a \textit{curiosity-encouraging} loss $\mathcal{L}_{\textrm{curious}}$, which minimizes the log likelihoods of actions in $A^{\textrm{nav}\otimes}$.
This loss prevents the agent from repeating past mistakes and motivates it to explore untried actions. 
The navigation loss is:
\begin{align}
    &\mathcal{L}_{\textrm{nav}}(s, \anavpol, \tnavpol) = \overbrace{-\log \hat{\pi}_{\textrm{nav}}(a^{\textrm{nav}\star} \mid o_s)}^{\mathcal{L}_{\textrm{NL}}(s, \anavpol, \tnavpol)}\\
    & \qquad\quad + \alpha \underbrace{\frac{1}{|A^{\textrm{nav}\otimes}|}\sum_{a \in A^{\textrm{nav}\otimes}} \log \hat{\pi}_{\textrm{nav}}(a \mid o_s)}_{\mathcal{L}_{\textrm{curious}}(s,\anavpol, \tnavpol) } \nonumber
\end{align} where $\alpha \in [0, \infty)$ is a weight hyperparameter.

\begin{figure*}[t!]
    \centering
    \includegraphics[width=0.85\textwidth]{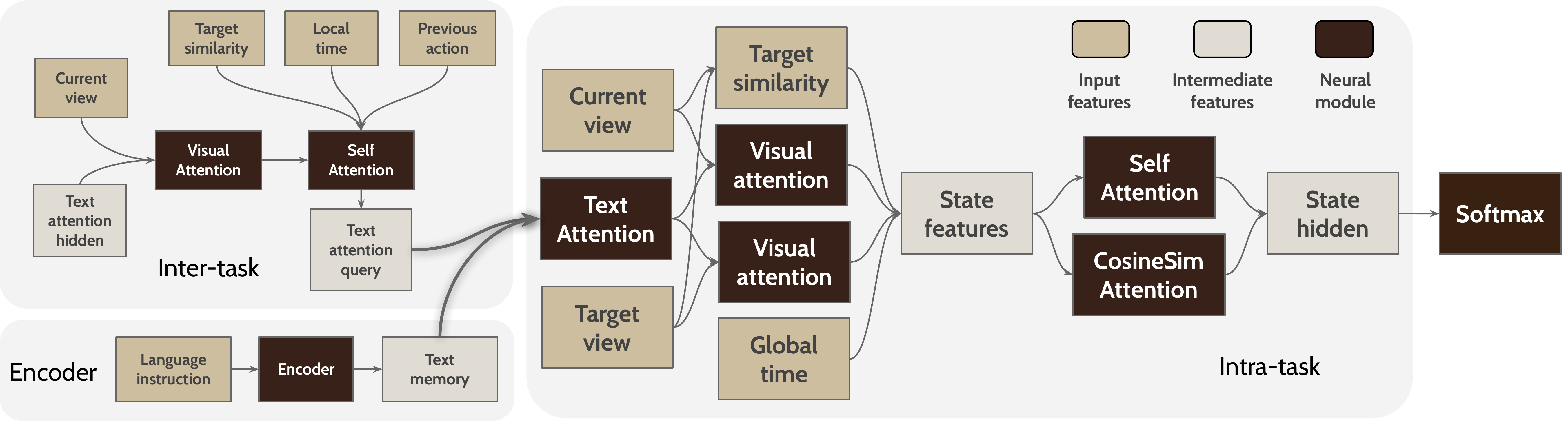}
    \caption{Our hierarchical recurrent model architecture (the navigation network). The help-request network is mostly similar except that the navigation action distribution is fed as an input to compute the ``state features".}
    \label{fig:my_label}
\end{figure*}

\paragraph{Retrospective Interpretable Help-Request Teacher.}
In deciding whether the agent should ask for help, the help-request teacher must consider the agent's future situations. 
Standard imitation learning algorithms (e.g., DAgger) employ an \emph{online} mode of interaction which queries the teacher at every time step. This mode of interaction is not suitable for our problem: the teacher must be able to predict the agent's future actions if it is queried when the episode is not finished. 
To overcome this challenge, we introduce a more efficient \emph{retrospective} mode of interaction, which waits until the agent completes an episode and queries the teacher for reference actions for \emph{all} time steps at once. 
With this approach, because the future actions at each time step are now fully observed, they can be taken into consideration when computing the reference action.
In fact, we prove that the retrospective teacher is optimal for teaching the agent to determine the \texttt{lost} condition, which is the only condition that requires knowing the agent's future. 

\begin{lemma} (proof in Appendix B) At any time step, the retrospective help-request teacher suggests the action that results in the agent getting closer to the target location in the future under its current navigation policy (if such an action exists). 
\end{lemma}

%\hal{i think i'd lead here with the fact that it's retrospective because it wants to be able to consider the agent's future behavior in deciding when to suggest it ask for help---this makes sense, i can't know if you need help if i don't have some model of what you're going to do next.}
%\kcomment{done}

To help the agent better justify its help-request decisions, we train a \emph{reason classifier} $\Phi$ to predict which conditions are satisfied. 
To train this classifier, the teacher provides a reason vector $\rho^{\star} \in \{0, 1\}^3$, where
$\rho_i^{\star} = 1$ indicates that the $i$-th condition is met. 
We formulate this prediction problem as multi-label binary classification and employ a binary logistic loss for each condition. 
Learning to predict the conditions helps the agent make more accurate and interpretable decisions. 
The help-request loss is:
\begin{align}
    &\mathcal{L}_{\textrm{ask}}(s, \aaskpol, \taskpol) = \overbrace{-\log \hat{\pi}_{\textrm{ask}}(a^{\textrm{ask}\star} \mid o_s)}^{\mathcal{L}^{\textrm{NL}}(s, \aaskpol, \taskpol)} \\
    & \qquad \underbrace{- \frac{1}{3}\sum_{i = 1}^3 \left[\rho_i^{\star}\log \hat{\rho}_i + (1 - \rho_i^{\star})\log (1 - \hat{\rho}_i) \right]}_{\mathcal{L}_{\textrm{reason}}(s, \aaskpol, \taskpol)} \nonumber
\end{align} where $\left( a^{\textrm{ask}\star}, \rho^{\star} \right) = \taskpol(s)$, and  $\hat{\rho} = \Phi(o_s)$ is the agent-estimated likelihoods of the conditions.

\section{Hierarchical Recurrent Architecture} \label{sec:architecture}
We model the navigation policy and the help-request policy as two separate neural networks.
The two networks have similar architectures, which consists of three main components: the \textit{text-encoding} component, the \textit{inter-task} component, and the \textit{intra-task} component (\autoref{fig:my_label}). 
We use self-attention instead of recurrent neural networks to better capture long-term dependency, and develop novel cosine-similarity attention and ResNet-based time-encoding.
Detail on the computations in each module is in the Appendix. 

The text-encoding component computes a text memory $M^{\textrm{text}}$, which stores the hidden representation of the current language instruction.
The inter-task module computes a vector $h^{\textrm{inter}}_t$ representing the state of the current task's execution.
During the episode, every time the current task is altered (due to the agent requesting help or departing a route), the agent re-encodes the new language instruction to generate a new text memory and resets the inter-task state to a zero vector.
The intra-task module computes a vector $h^{\textrm{intra}}_t$ representing the state of the entire episode. To compute this state, we first calculate $\bar{h}^{\textrm{intra}}_t$, a tentative current state, and $\tilde{h}^{\textrm{intra}}_t$, a weighted combination of the past states at nearly identical situations. 
$h^{\textrm{intra}}_t$ is computed as: % by subtracting scaled version of $\tilde{h}^{\textrm{intra}}_t$ from $\bar{h}^{\textrm{intra}}_t$
\begin{align}
    h^{\textrm{intra}}_t &= \bar{h}^{\textrm{intra}}_t - \beta \cdot \tilde{h}^{\textrm{intra}}_t  \label{eqn:state_dissimilar} \\
    \beta &= \sigma(W_{\textrm{gate}} \cdot [\bar{h}^{\textrm{intra}}_t; \tilde{h}^{\textrm{intra}}_t]) 
\end{align}
\autoref{eqn:state_dissimilar} creates an context-sensitive dissimilarity between the current state and the past states at nearly identical situations.
The scale vector $\beta$ determines how large the dissimilarity is based on the inputs. 
This formulation incorporates past related information into the current state, thus enables the agent to optimize the curiosity-encouraging loss effectively. 
Finally, $h^{\textrm{intra}}_t$ is passed through a softmax layer to produce an action distribution.

\nocite{vaswani2017attention}
\nocite{he2016deep}
\nocite{russakovsky2015imagenet}
\nocite{fried2018speaker}
\nocite{lei2016layer}

\section{Experimental Setup} \label{sec:exp}
\begin{table}[t!]
    \small
    \centering
    \setlength{\tabcolsep}{2.5pt}
	\begin{tabular}{lccc}
		\toprule
        Split & Environments & Tasks & \system Instructions \\
		\cmidrule(lr{0.5em}){1-1} \cmidrule(lr{0.5em}){2-2} \cmidrule(lr{0.5em}){3-3} \cmidrule(lr{0.5em}){4-4}
		Train & 51 & 82,484 & 8,586 \\
		Val SeenEnv & 51 & ~~5,001 & 4,287 \\
		Val UnseenAll & ~~7 & ~~5,017 & 2,103 \\
		Test SeenEnv & 51 & ~~5,004 & 4,287 \\
		Test Unseen & 10 & ~~5,012 & 2,331 \\
        \bottomrule
	\end{tabular}
	\smallskip
	\caption{Data split.}
	\label{tab:data}
\end{table}

\begin{table*}[t!]
    \small
    \centering
    \setlength{\tabcolsep}{3pt}
	\begin{tabular}{l@{~~~~~~~~~}ccccp{0.5cm}cccc}
		\toprule[1pt]
	    & \multicolumn{4}{c}{\unseenLang} && \multicolumn{4}{c}{\unseenAll} \\
	    \cmidrule[.002em](lr{0.4em}){2-5}  \cmidrule[.002em](lr{0.4em}){7-10}
	    \multicolumn{1}{l}{Agent} & SR $\uparrow$ & SPL $\uparrow$ & Nav. $\downarrow$ & Requests/&
	    & SR $\uparrow$ & SPL $\uparrow$ & Nav.  $\downarrow$ & Requests/  \\
	    & (\%) & (\%) & Err. (m) & task $\downarrow$&
	    & (\%) & (\%) & Err. (m) & task $\downarrow$\\
	    \specialrule{.002em}{0.3em}{-0.8em} \\
	    \textbf{Non-learning agents} \\
	    %\addlinespace[0.4em]
	            \randomWalk & ~~~~0.54 & ~~~~0.33 & 15.38 & 0.0 && ~~~~0.46 & ~~~~0.23 & ~~15.34 & 0.0 \\ 
	               \forward & ~~~~5.98 & ~~~~4.19 & 14.61 & 0.0 && ~~~~6.36 & ~~~~4.78 & ~~13.81 & 0.0 \\ 
	    \specialrule{.002em}{0.3em}{-0.8em} \\
	    \textbf{Learning agents} \\
	    %\addlinespace[0.4em]
	    No assistance & ~~17.21 & ~~13.76 & 11.48 & 0.0 && ~~~~8.10 & ~~~~4.23 & 13.22 & 0.0 \\ 
	    Learn to interpret assistance (ours) & ~~\textbf{88.37} & ~~\textbf{63.92} & ~~\textbf{1.33} & \textbf{2.9} && ~~\textbf{47.45} & ~~\textbf{25.50} & ~~\textbf{7.67} & \textbf{5.8} \\ 
	    \specialrule{.002em}{0.3em}{-0.8em} \\
	    \textbf{Skylines} \\
	    %\addlinespace[0.4em]
	    \shortest & 100.00 & 100.00 & ~~0.00 & 0.0 && 100.00 & 100.00 & ~~0.00 & 0.0\\ 
	    Perfect assistance interpretation & ~~90.99 & ~~68.87 & ~~0.91 & 2.5 && ~~83.56 & ~~56.88 & ~~1.83 & 3.2 \\
        \bottomrule[1pt] 
	\end{tabular}
	\caption{Results on test splits. The agent with ``perfect assistance interpretation" uses the teacher navigation policy ($\tnavpol$) to make decisions when executing a subtask from \system. Results of our final system are in bold.}
	\label{tab:main}
\end{table*}

\paragraph{Dataset.} We generate a dataset of object-finding tasks in the \problem environments to train and evaluate our agent.
\autoref{tab:data} summarizes the dataset split.
Our dataset features 289 object types; the language instruction vocabulary contains 2,332 words. 
The numbers of locations on the shortest paths to the requested objects are restricted to be between 5 and 15. 
With an average edge length of 2.25 meters, the agent has to travel about 9 to 32 meters to reach its goals.
We evaluate the agent in environments that are seen during training (\unseenLang), and in environments that are not seen  (\unseenAll).
Even in the case of \unseenLang, the tasks and the \system language instructions given during evaluation were never given in the same environments during training.

\paragraph{Hyperparameters.} See Appendix.

\paragraph{Baselines and Skylines.} We compare our agent against the following \emph{non-learning} agents:
\begin{enumerate*} %[label={$\diamond$},nolistsep]
    \item \textsc{Shortest}: uses the navigation teacher policy to make decisions (this is a skyline);
    \item \textsc{RandomWalk}: randomly chooses a navigation action at every time step;
    \item \textsc{Forward10}: navigates to the next location closest to the center of the current view to advance for 10 time steps. 
\end{enumerate*}
We compare our learned help-request policy with the following \emph{heuristics}:
\begin{enumerate*} %[label={$\diamond$},nolistsep]
    \item \textsc{NoAsk}: does not request help;
    \item \textsc{RandomAsk}: randomly chooses to request help with a probabilty of 0.2, which is the average help-request ratio of our learned agent;
    \item \textsc{AskEvery5}: requests help as soon as walking at least 5 time steps.
\end{enumerate*}

\paragraph{Evaluation metrics.} Our main metrics are: 
\emph{success rate} (SR), the fraction of examples on which the agent successfully solves the task;
\emph{navigation error}, the average (shortest-path) distance between the agent's final location and the nearest goal from that location;
and \emph{SPL} \citep{anderson2018evaluation}, which weights task success rate by travel distance as follows:
\begin{align}
    \text{SPL} = \frac{1}{N} \sum_{i = 1}^N S_i \frac{L_i}{\max(P_i, L_i)} 
\end{align} where $N$ is the number of tasks, $S_i$ indicates whether task $i$ is successful, $P_i$ is the agent's travel distance, and $L_i$ is the shortest-path distance to the goal nearest to the agent's final location. 
%The success radius $\epsilon$ is 2 meters in all experiments; there is always at least one such node.

\section{Results} \label{sec:results}

\paragraph{Main results.}  From \autoref{tab:main}, we see that our problem is challenging: simple heuristic-based baselines such as \textsc{RandomWalk} and \textsc{Forward10} attain success rates less than 7\%. 
An agent that learns to accomplish tasks without additional assistance from \system succeeds only 17.21\% of the time on \testSet \unseenLang, and 8.10\% on \testSet \unseenAll. 
Leveraging help from \system dramatically boosts the success rate by 71.16\% on \testSet \unseenLang and by 39.35\% on \testSet \unseenAll over not requesting help. Given the small size of our dataset (e.g., the agent has fewer than 9,000 subtask instructions to learn from), it is encouraging that our agent is successful in nearly half of its tasks.
On average, the agent takes paths that are 1.38 and 1.86 times longer than the optimal paths on \testSet \unseenLang and \testSet \unseenAll, respectively. 
In unseen environments, it issues on average twice as many requests to as it does in seen environments. To understand how well the agent interprets the \system instructions, we also provide results where our agent uses the optimal navigation policy to make decisions while executing subtasks. 
The large gaps on \testSet \unseenLang indicate there is still much room for improvement in the future, purely in learning to execute language instructions. 

\begin{table}[t!]
    \small
    \centering
	\begin{tabular}{lcc}
		\toprule[1pt]
		 \multicolumn{1}{l}{Assistance type}& \unseenLang & \unseenAll \\
		 \cmidrule(lr){1-1} \cmidrule(lr){2-2} \cmidrule(lr){3-3}
		Target image only & 84.95 & 31.88 \\
		+ Language instruction & \textbf{88.37} & \textbf{47.45}\\
        \bottomrule[1pt] 
	\end{tabular}
	\caption{Success rates (\%) of agents on test splits with different types of assistance.}
	\label{tab:assistance_type}
\end{table}
\paragraph{Does understanding language improve generalizability?}
Our agent is assisted with both language and visual instructions; similar to \citet{thomason2018shifting}, we disentangle the usefulness two these two modes of assistance.
%When our agent only has access to images, performance degrades from $88.37$ to $84.95$ on seen environments, and from $47.45$ to $31.88$ on unseen environments.
As seen in \autoref{tab:assistance_type}, the improvement from language on \testSet \unseenAll (+15.17\%) is substantially more than that on \testSet \unseenLang (+3.42\%), largely the agent can simply memorize the seen environments. 
This confirms that understanding language-based assistance effectively enhances the agent's capability of accomplishing tasks in novel environments.  

\begin{table}[t!]
    \small
    \centering
    \setlength{\tabcolsep}{4pt}
	\begin{tabular}{lccp{0cm}cc}
		\toprule[1pt]
		 & 
		 \multicolumn{2}{c}{\unseenLang} && 
		 \multicolumn{2}{c}{\unseenAll} \\
		 \cmidrule[.002em](lr){2-3}  \cmidrule[.002em](lr){5-6}
		 \multicolumn{1}{l}{$\aaskpol$} & SR $\uparrow$ & Requests/ && SR $\uparrow$ & Requests/ \\
		 & (\%) & task $\downarrow$ && (\%) & task $\downarrow$ \\
		 \cmidrule[.002em](lr){1-1}  \cmidrule[.002em](lr){2-3}  \cmidrule[.002em](lr){5-6}
		\noAsk & ~17.21 & 0.0 && ~~8.10 & 0.0 \\
		\randomAsk & ~82.71 & 4.3 && 37.05 & 6.8 \\
		\textsc{AskEvery5} & 87.39 & 3.4 && 34.42 & 7.1 \\
		Learned (ours) & \textbf{88.37} & \textbf{2.9} && \textbf{47.45} & \textbf{5.8} \\
        \bottomrule[1pt] 
	\end{tabular}
	\caption{Success rates (\%) of different help-request policies on test splits.}
	\label{tab:compare_askpol}
\end{table}

\paragraph{Is learning to request help effective?} \autoref{tab:compare_askpol} compares our learned help-request policies with baselines. 
We find that \textsc{AskEvery5} provides a surprisingly strong baseline for this problem, leading to an improvement of +26.32\% over not requesting help on \testSet \unseenAll. 
Nevertheless, our learned policy, with the ability to predict the future and access to the agent's uncertainty, outperforms all baselines by at least 10.40\% in success rate on \testSet \unseenAll, while making less help requests.
The small gap between the learned policy and \textsc{AskEvery5} on \testSet \unseenAll is expected because, on this split, the performance is mostly determined by the model's memorizing capability and is mostly insensitive to the help-request strategy.

\begin{table}[t!]
    \small
    \centering
    \setlength{\tabcolsep}{2.4pt}
	\begin{tabular}{lccc}
		\toprule[1pt]
		 & SR $\uparrow$ & Nav. mistake $\downarrow$ & Help-request $\downarrow$ \\
		 Model & (\%) & repeat (\%) & repeat (\%)
		 \\ \midrule[0.002em] 
		\textsc{LSTM-EncDec} & 19.25 & 31.09 & 49.37 \\ 
		Our model ($\alpha = 0$) & 43.12 & 25.00 & 40.17 \\
		%Our model ($\alpha = 0$) & 39.29 & 20.75 & 40.77 \\
		Our model ($\alpha = 1$) & \textbf{47.45}& \textbf{17.85} & \textbf{21.10} \\ 
		%Our model ($\alpha = 1$) & \textbf{45.64}& \textbf{19.24} & \textbf{21.14} \\ 
        \bottomrule[1pt] 
	\end{tabular}
	\caption{Results on \testSet \unseenAll of our model, trained with and without curiosity-encouraging loss, and an LSTM-based encoder-decoder model (both models have about 15M parameters). ``Navigation mistake repeat" is the fraction of time steps on which the agent repeats a non-optimal navigation action at a previously visited location while executing the same task. 
	``Help-request repeat" is the fraction of help requests made at a previously visited location while executing the same task.}
	\label{tab:proposed_method}
\end{table}

\paragraph{Is proposed model architecture effective?}
We implement an LSTM-based encoder-decoder model that is based on the architecture proposed by \citep{wang2018rcm-sil}. 
To incorporate the target image, we add an attention layer that uses the image's vector set as the attention memory. 
We train this model with imitation learning using the standard negative log likelihood loss (\autoref{eqn:nl_loss}), without the curiosity-encouraging and reason-prediction losses. 
As seen in \autoref{tab:proposed_method}, our hierarchical recurrent model outperforms this model by a large margin on \testSet \unseenAll (+28.2\%).

\paragraph{Does the proposed imitation learning algorithm achieve its goals?}
The curiosity-encouraging training objective is proposed to prevent the agent from making the same mistakes at previously encountered situations. 
\autoref{tab:proposed_method} shows that training with the curiosity-encouraging objective reduces the chance of the agent looping and making the same decisions repeatedly.
As a result, its success rate is greatly boosted (+4.33\% on \testSet \unseenAll) over no curiosity-encouraging.

\section{Conclusion} \label{sec:conclusion}
In this work, we present a photo-realistic simulator that mimics primary characteristics of real-life human assistance. 
We develop effective imitation learning techniques for learning to request and interpret the simulated assistance, coupled with a hierarchical neural network model for representing subtasks.
Future work aims to provide more natural, linguistically realistic interaction between the agent and humans (e.g., providing the agent the ability ask a natural \emph{question} rather than just signal for help), and to establish a theoretical framework for modeling human assistance.
We are also exploring ways to deploy and evaluate our methods on real-world platforms. 

\bibliography{journal-full,emnlp-ijcnlp-2019}
\bibliographystyle{acl_natbib}

\newpage
\setcounter{section}{0}
\title{Appendix to \\ Help, Anna! Visual Navigation with Natural Multimodal Assistance via Retrospective Curiosity-Encouraging Imitation Learning}

\author{}
\appendix
\maketitle

\setcounter{lemma}{0}

\section*{Acknowledgement}

We would like to thank Debadeepta Dey for helpful discussions. We thank Jesse Thomason for the thorough review and suggestions on related work and future directions. Many thanks to the reviewers for their meticulous and insightful comments.
Special thanks to Reviewer 2 for addressing our author response in detail, and to Reviewer 3 for the helpful suggestions and the careful typing-error correction.

\section{Proof of Lemma 1}

\begin{lemma} To guide the agent between any two locations using $O(\log N)$ instructions, we need to collect instructions for $\Theta(N \log N)$ location pairs.
\end{lemma}

\begin{proof}
Consider a spanning tree of the environment graph that is rooted at $v_r$. For each node $v$, let $d(v)$ be the distance from $v$ to $v_r$. For $i = 0 \cdots \floor{\log_2 d(v)} - 1$, collect an instruction for the path from $v$ to its $2^i$-th ancestor, and an instruction for the reverse path.
In total, no more than $2 \cdot N \log_2 N = \Theta(N \log N)$ instructions are collected.

A language-assisted path is a path that is associated with a natural language instruction. 
It is easy to show that, under this construction, $O(\log N)$ language-assisted paths are sufficient to traverse between $v$ and any ancestor of $v$ as follows.
For any two arbitrary nodes $u$ and $v$, let $w$ be their least common ancestor. 
To traverse from $u$ to $v$, we first go from $u$ to $w$, then from $w$ to $v$. 
The total number of language-assisted path is still $O(\log N)$.
\end{proof}

\section{Proof of Lemma 2}

\begin{lemma} At any time step, the retrospective help-request teacher suggests the action that results in the agent getting closer to the target location in the future under its current navigation policy (if such an action exists). 
\end{lemma}

The retrospective mode of interaction allows the teacher to observe the agent's full trajectory when computing the reference actions. Note that this trajectory is generated by the agent's own navigation policy ($\anavpol$) because, during training, we use the agent's own policies to act. During a time step, after observing the future trajectory, the teacher determines whether the \texttt{lost} condition is satisfied or not, i.e. whether the agent will get closer to the target location or not. 
\begin{enumerate}[nolistsep]
    \item If condition is \emph{not} met, the teacher suggests the \texttt{do\_nothing} action: it is observed that the agent will get closer the target location without requesting help by following the current navigation policy.
    \item If condition is met, the teacher suggests the \texttt{request\_help} action. First of all, in this case, it is observed that the \texttt{do\_nothing} action will lead to no progress toward the target location. If the \texttt{request\_help} action will also lead to no progress, then it does not matter which action the teacher suggests. In contrast, if the \texttt{request\_help} action will help the agent get closer to the target location, the teacher has suggested the desired action. 
\end{enumerate}

\section{Features}

At time step $t$, the model makes use of the following features:
\begin{itemize}[label={$\diamond$},nolistsep]
    \item $I_t^{\textrm{cur}}$: set of visual feature vectors representing the current panoramic view;
    \item $I_t^{\textrm{tgt}}$: set of visual feature vectors representing the target location's panoramic view;
    \item $a_{t - 1}$: embedding of the last action;
    \item $\delta_t$: a vector representing how similar the target view is to the current view;
    \item $\eta^{\textrm{loc}}_t$: local-time embedding encoding the number of time steps since the last time the inter-task module was reset;
    \item $\eta^{\textrm{glob}}_t$: global-time embedding encoding the number time steps since the beginning of the episode;
    \item $P^{\textrm{nav}}_t$: the navigation action distribution output by $\anavpol$. Each action corresponding to a view angle is mapped to a static index to ensure that the order of the actions is independent of the view angle. This feature is only used by the help-request network. 
\end{itemize}

\paragraph{Visual features.}
To embed the first-person view images, we use the visual feature vectors provided \citet{anderson2018vision}, which are extracted from a ResNet-152 \citep{he2016deep} pretrained on ImageNet \citep{russakovsky2015imagenet}. 
Following the Speaker-Follower model \citep{fried2018speaker}, at time step $t$, we provide the agent with a feature set $I_t^{\textrm{cur}}$ representing the current panoramic view. 
The set consists of visual feature vectors that represent all 36 possible first-person view angles (12 headings $\times$ 3 elevations). 
Similarly, the panoramic view at the target location is given by a feature set $I_t^{\textrm{tgt}}$.
Each next location is associated with a view angle whose center is closest to it (in angular distance).
The embedding of a navigation action $(v, \Delta\psi, \Delta\omega)$ is constructed by concatenating the feature vector of the corresponding view and an orientation feature vector $[\sin\Delta\psi; \cos\Delta\psi;\sin\Delta\omega;\sin\Delta\omega]$ where $\Delta\psi$ and $\Delta\omega$ are the camera angle change needed to shift from the current view to the view associated with $v$.
The \texttt{stop} action is mapped to a zero vector. 
The action embedding set $E_t^{\textrm{nav}}$ contains embeddings of all navigation actions at time step $t$. 

\paragraph{Target similarity features.} The vector $\delta$ represents the similarity between $I^{\textrm{cur}}$ and $I^{\textrm{tgt}}$. 
To compute this vector, we first construct a matrix $C$, where $C_{i,j}$ is the cosine similarity between $I^{\textrm{cur}}_i$ and $I^{\textrm{tgt}}_j$. 
$\delta_t$ is then obtained by taking the maximum values of the rows of $C$.
The intuition is, for each current view angle, we find the most similar target view angle. 
If the current view perfectly matches with the target view, $\delta_t$ is a vector of ones. 

\paragraph{Time features.}
The local-time embedding is computed by a residual neural network  \citep{he2016deep} that learns a time-incrementing operator
\begin{align}
    \eta^{\textrm{loc}}_t = \textsc{IncTime}\left( \eta^{\textrm{loc}}_{t - 1}\right) 
\end{align} where $\textsc{IncTime}(x) = \textsc{LayerNorm}(x + \textsc{ReLu}(\textsc{Linear}(x)))$. 
The global-time embedding is computed similarly but also incorporates the current local-time
\begin{align}
    \eta^{\textrm{glob}}_t = \textsc{IncTime}\left( \left[\eta^{\textrm{loc}}_{t};\eta^{\textrm{glob}}_{t - 1} \right] \right) 
\end{align} The linear layers of the local and global time modules do not share parameters.
Our learned time features generalizes to previously unseen numbers of steps.
They allow us to evaluate the agent on longer episodes than during training, significantly reducing training cost.
We use the sinusoid encoding \citep{vaswani2017attention} for the text-encoding module, and the ResNet-based encoding for the decoder modules.
In our preliminary experiments, we also experimented using the sinusoid encoding in all modules but doing so significantly degraded success rate. 

\section{Model}

\paragraph{Modules.} The building blocks of our architecture are the Transformer modules \citep{vaswani2017attention}
\begin{itemize}[label={$\diamond$},nolistsep]
    \item \textsc{TransEncoder}$(l)$ is a Transformer-style encoder, which generates a set of feature vectors representing an input sequence $l$;
    \item $\textsc{MultiAttend}(q, K, V)$ is a multi-headed attention layer that takes as input a query vector $q$, a set of key vectors $V$, and a set of value vectors $V$. The output is added a residual connection \citep{he2016deep} and is layer-normalized \citep{lei2016layer}.
    \item $\textsc{FFN}(x)$ is a Transformer-style feed-forward layer, which consists of a regular feed-forward layer with a hidden layer and the \textsc{ReLU} activation function, followed by a residual connection and layer normalization. The hidden size of the feed-forward layer is four times larger than the input/output size.
\end{itemize}

In addition, we devise the following new attention modules
\begin{itemize}[label={$\diamond$},nolistsep]
    \item \textsc{SelfAttend}$(q, K, V)$ implements a multi-headed attention layer internally. It calculates an output $h = \textsc{MultiAttend}(q, K, V)$. After that, the input $q$ and output $h$ are appended to $K$ and $V$, respectively. This module is different from the Transformer's self-attention in that the keys and values are distinct.
    \item \textsc{SimAttend}$(q, K, V)$ computes a weighted value $h = \sum_i \tilde{a}_i V_i$ where each weight $\tilde{a}_i$ is defined as 
    \begin{align}
        \tilde{a}_i &= \mathbb{I}\{ a_i > 0.9 \} \frac{a_i}{\sum_j a_j} \\
        a_i &= \textsc{CosineSimilarity}(q, K_i) 
    \end{align} where $\textsc{CosineSimilarity}(.,.)$ returns the cosine similarity between two vectors, and $\mathbb{I}\{.\}$ is an indicator function. Intuitively, this module finds keys that are nearly identical to the query and returns the weighted average of values corresponding to those keys. We use this module to obtain a representation of related past, which is crucial to enforcing curiosity-encouraging training.  
\end{itemize}

We now describe the navigation network in detail. For notation brevity, we omit the ${}^\textrm{nav}$ superscripts in all variables.

\begin{table}[t!]
        \small
        \centering
        \begin{tabular}{lc}
        \toprule
        \multicolumn{1}{c}{Hyperparameter} & Value \\ \midrule
        \multicolumn{2}{c}{Common} \\
        Hidden size & 256 \\
        Navigation action embedding size & 256 \\ 
        Help-requesting action embedding size & 128 \\ 
        Word embedding size & 256 \\
        Number of self-attention heads & 8 \\
        Number of instruction-attention heads & 8 \\
        ResNet-extracted visual feature size & 2048 \\
        \midrule
        \multicolumn{2}{c}{Help-request teacher} \\
        Uncertainty threshold ($\gamma$) & 0.25 \\
         \midrule
        \multicolumn{2}{c}{Training} \\
        Optimizer & Adam \\
        Number of training iterations & $3 \times 10^4$ \\
        Learning rate & $10^{-4}$ \\ 
        Training batch size & 32 \\
        Dropout ratio & 0.3 \\
        Training time steps & 20 \\
        Maximum instruction length & 50 \\
        Curiosity-encouraging weight ($\alpha$) & 1.0$^{(*)}$ \\ \midrule
        \multicolumn{2}{c}{Evaluation} \\
        Success radius ($\epsilon_{\textrm{success}}$) & 2 meters \\
        Attention radius ($\epsilon_{\textrm{attn}}$) & 2 meters \\
        Evaluation time steps & 50 \\
        Evaluation batch size & 32 \\
        \bottomrule
       \end{tabular}
     \caption{Hyperparameters. (*) Training collapses when using $\alpha = 1$ to train the agent with the help-request policy baselines (\textsc{NoAsk}, \textsc{RandomAsk}, \textsc{AskEvery5}). Instead, we use $\alpha = 0.5$ in those experiments. }
     \label{tab:hyper}
\end{table}

\paragraph{Text-encoding component.} The agent maintains a text memory $M^{\textrm{text}}$, which stores the hidden representation of the current language instruction $l_t$. 
At the beginning of time, the agent encodes the task request to generate an initial text memory.
During time step $t$, if the current task is altered (due to the agent requesting help or departing a route), the agent encodes the new language instruction to generate a new text memory
\begin{align}
    M^{\textrm{text}} =
    \textsc{TransEncoder}(l_t) \\ \text{if } l_t \neq l_{t - 1} \ \text{or } t = 0 \nonumber
\end{align}

\paragraph{Inter-task component.} The inter-task module computes a vector $h^{\textrm{inter}}_t$ representing the state of the current task's execution.
This state and the local time embedding are reset to zero vectors every time the agent switches task.
Otherwise, a new state is computed as follows
\begin{align}
    h^{\textrm{inter}}_t &= 
    \textsc{SelfAttend}(q_t^{\textrm{inter}}, M^{\textrm{inter}}_{\textrm{in}}, M^{\textrm{inter}}_{\textrm{out}}) \\
    q_t^{\textrm{inter}} &= W_{\textrm{inter}} [c_t^{\textrm{inter}}; a_{t - 1}; \delta_t] + \eta^{\textrm{loc}}_t \label{eqn:inter_input} \\
    c_t^{\textrm{inter}} &= \textsc{DotAttend}(h^{\textrm{inter}}_{t - 1}, I_t^{\textrm{cur}})
\end{align} where $M^{\textrm{inter}}_{\textrm{in}} = \{ q_0^{\textrm{inter}}, \cdots, q_{t - 1}^{\textrm{inter}} \}$ and $M^{\textrm{inter}}_{\textrm{in}} = \{ h_0^{\textrm{inter}}, \cdots, h_{t - 1}^{\textrm{inter}} \}$ are the input and output inter-task memories, and $\textsc{DotAttend}(q, M)$ is the dot-product attention \citep{luong2015effective}.

Next, $h^{\textrm{inter}}_t$ is used to select which part of the language instruction should be interpreted in this step
\begin{align}
c^{\textrm{text}}_t &= \textsc{FFN}\left(\tilde{c}^{\textrm{text}}\right) \\
\tilde{c}^{\textrm{text}} &=    \textsc{MultiAttend}(h^{\textrm{inter}}_t, M^{\textrm{text}}_t, M^{\textrm{text}}_t)
\end{align}

Finally, $c^{\textrm{text}}_t$ is used to select which areas of the current view and target view the agent should focus on
\begin{align}
    c^{\textrm{cur}}_t &= \textsc{DotAttend}(c^{\textrm{text}}_{t}, I_t^{\textrm{cur}}) \\
    c^{\textrm{tgt}}_t &= \textsc{DotAttend}(c^{\textrm{text}}_{t}, I_t^{\textrm{tgt}})
\end{align}

\paragraph{Intra-task component.} The inter-task module computes a vector $h^{\textrm{intra}}_t$ representing the state of the entire episode. To compute this state, we first calculate $\bar{h}^{\textrm{intra}}_t$, a tentative current state, and $\tilde{h}^{\textrm{intra}}_t$, a weighted combination of the past states at nearly identical situations 
\begin{align}
    \bar{h}^{\textrm{intra}}_t &= 
    \textsc{FFN}\left(\textsc{SelfAttend}(q_t^{\textrm{intra}}, M^{\textrm{intra}}_{\textrm{in}}, M^{\textrm{intra}}_{\textrm{out}})\right) \\
    \tilde{h}^{\textrm{intra}}_t &= 
    \textsc{SimAttend}(q_t^{\textrm{intra}},
    M^{\textrm{intra}}_{\textrm{in}}, M^{\textrm{intra}}_{\textrm{out}}) \\
    q_t^{\textrm{intra}} &= W_{\textrm{intra}}\left[c^{\textrm{text}}_t; c^{\textrm{cur}}_t;  c^{\textrm{tgt}}_t; \delta_t \right] + \eta^{\textrm{glob}}_t \label{eqn:intra_input}
\end{align} where $M^{\textrm{intra}}_{\textrm{in}} = \{ q_0^{\textrm{intra}}, \cdots, q_{t - 1}^{\textrm{intra}} \}$ and $M^{\textrm{intra}}_{\textrm{in}} = \{ h_0^{\textrm{intra}}, \cdots, h_{t - 1}^{\textrm{intra}} \}$ are the input and output intra-task memories.
The final state is determined by subtracting a scaled version of $\tilde{h}^{\textrm{intra}}_t$ from $\bar{h}^{\textrm{intra}}_t$
\begin{align}
    h^{\textrm{intra}}_t &= \bar{h}^{\textrm{intra}}_t - \beta \cdot \tilde{h}^{\textrm{intra}}_t  \label{eqn:state_dissimilar} \\
    \beta &= \sigma(W_{\textrm{gate}} \cdot [\bar{h}^{\textrm{intra}}_t; \tilde{h}^{\textrm{intra}}_t]) 
\end{align} 
Finally, the navigation action distribution is computed as follows
\begin{align}
    P^{\textrm{nav}}_{t} &= \textsc{Softmax}(W^{\textrm{out}} ~ h^{\textrm{intra}}_t ~ W^{\textrm{act}} ~ E_t^{\text{nav}})) 
\end{align} where $E_t^{\text{nav}}$ is a matrix containing embeddings of the navigation actions. 
The computations in the help-request network is almost identical except that (a) the navigation action distribution $P^{\textrm{nav}}_t$ is fed as an extra input to the intra-task components ( \ref{eqn:intra_input}), and (b) the help-request and reason distributions are calculated as follows
\begin{align}
    P^{\textrm{ask}}_{t} &= \textsc{Softmax}(E_t^{\text{ask}}E_t^{\text{reason}} h^{\textrm{ask,intra}}_t) \\
    P^{\textrm{reason}}_{t} &= \textsc{Softmax}(E_t^{\text{reason}} h^{\textrm{ask,intra}}_t) 
\end{align} where $E_t^{\text{ask}}$ contains embeddings of the help-request actions and $E_t^{\text{reason}}$ contains embeddings of the reason labels. 

\begin{table*}[t!]
    \small
    \centering
    \setlength{\tabcolsep}{3pt}
	\begin{tabular}{l@{~~~~~~~~~}ccccp{0.5cm}cccc}
		\toprule[1pt]
	    & \multicolumn{4}{c}{\valSet \unseenLang} && \multicolumn{4}{c}{\valSet \unseenAll} \\
	    \cmidrule[.002em](lr{0.4em}){2-5}  \cmidrule[.002em](lr{0.4em}){7-10}
	    \multicolumn{1}{l}{Agent} & SR $\uparrow$ & SPL $\uparrow$ & Nav. $\downarrow$ & Requests/&
	    & SR $\uparrow$ & SPL $\uparrow$ & Nav.  $\downarrow$ & Requests/  \\
	    & (\%) & (\%) & Err. (m) & task $\downarrow$&
	    & (\%) & (\%) & Err. (m) & task $\downarrow$\\
	    \specialrule{.002em}{0.3em}{-0.8em} \\
	    %\addlinespace[0.4em]
	    Final agent & ~~87.24 & ~~63.02 & \textbf{1.21} & 2.9 && ~~45.64 & ~~22.68 & ~~7.72 & \textbf{5.9} \\ 
	    No inter-task reset & ~~83.62 & ~~60.44 & 1.44 & 3.2 && ~~40.60 & ~~19.89 & ~~8.26 & 7.3 \\ 
	    No condition prediction & ~~72.33 & ~~47.10 & 2.64 & 4.7 && ~~\textbf{47.88} & ~~\textbf{24.68} & ~~\textbf{6.61} & 7.9 \\ 
	    No cosine-similarity attention ($\beta = \vec 0$) & ~~83.08 & ~~59.52 & 1.68 & 3.0 && ~~43.69 & ~~22.44 & ~~7.94 & 6.9 \\
	    No curiosity-encouraging loss ($\alpha = 0$) & ~~\textbf{87.36} & ~~\textbf{70.18} & 1.25 & \textbf{2.0} && ~~39.23 & ~~20.78 & ~~8.88 & 7.5 \\
        \bottomrule[1pt] 
	\end{tabular}
	\caption{Ablation study on our proposed techniques. Models are evaluated on the validation splits and with a batch size of 32. Best results in each column are in bold.}
	\label{tab:ablation}
\end{table*}

\section{Hyperparameters}

\autoref{tab:hyper} summarizes all training and evaluation hyperparameters. Training our agent takes approximately 9 hours on a single Titan Xp GPU. 

\section{Analysis}
\label{sec:analysis}

\begin{figure}[t!]
     \centering
     \begin{subfigure}[b]{0.37\linewidth}
         \centering
         \includegraphics[width=\textwidth]{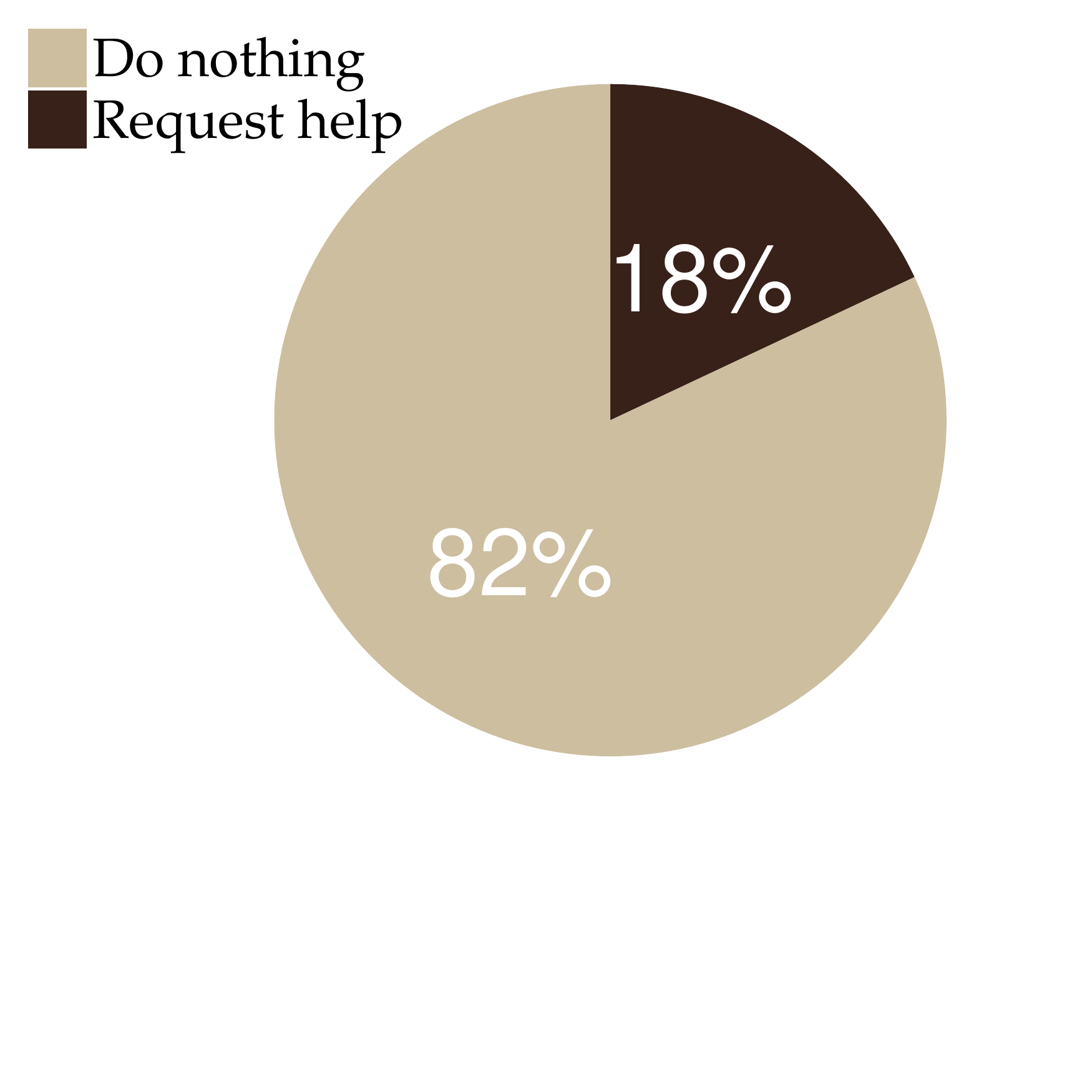}
         \caption{}
         \label{fig:ask_dist}
     \end{subfigure}
     \hfill
     \begin{subfigure}[b]{0.6\linewidth}
         \centering
         \includegraphics[width=\textwidth]{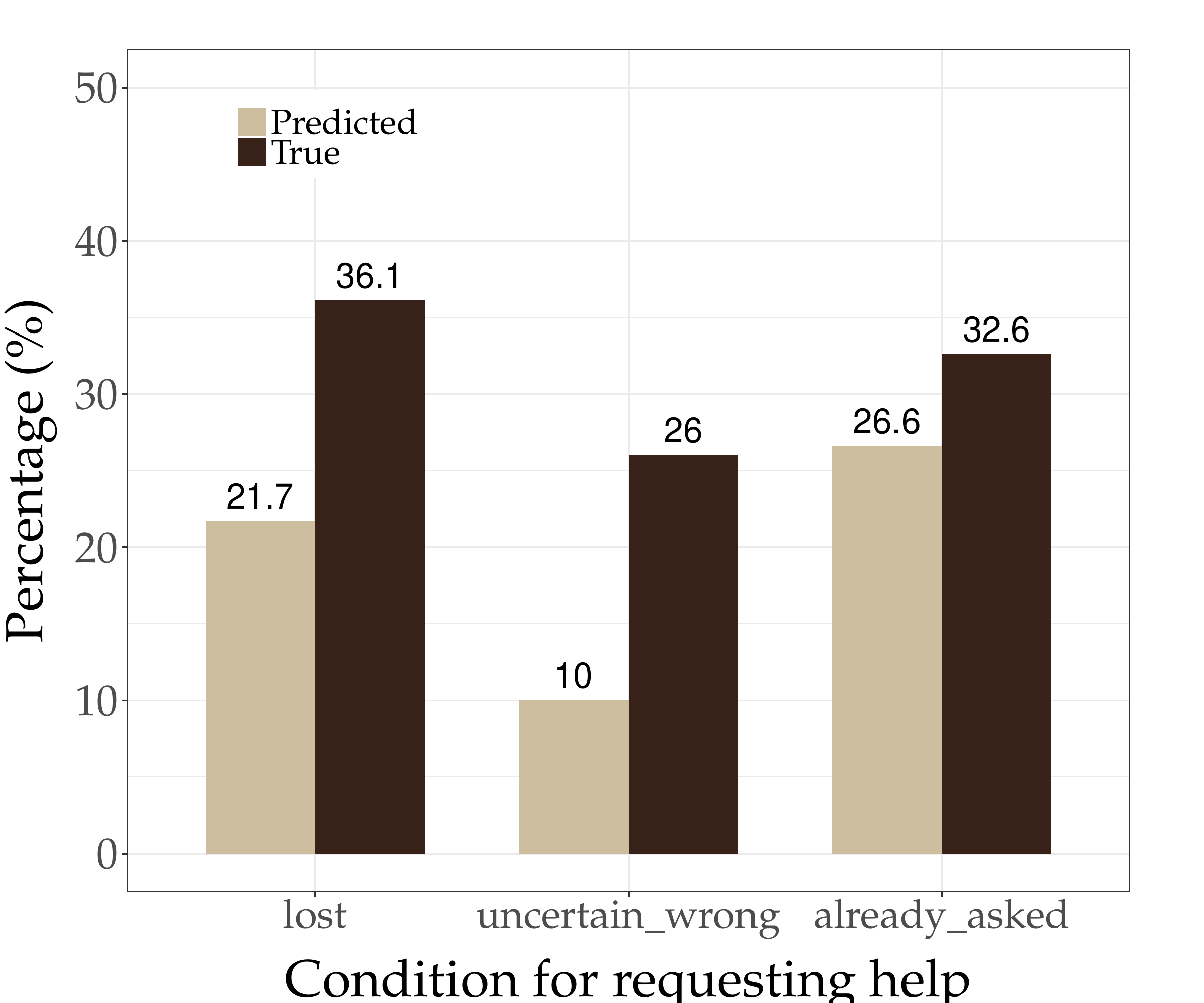}
         \caption{}
         \label{fig:reason_dist}
     \end{subfigure}
     \caption{Help-request behavior on \testSet \unseenAll: (a) fraction of time steps where the agent requests help and (b) predicted and true condition distributions. The \texttt{already\_asked} condition is the negation of the \texttt{never\_asked} condition.}
     \label{fig:three graphs}
\end{figure}

\begin{figure}[t!]
    \centering
    \includegraphics[width=\linewidth]{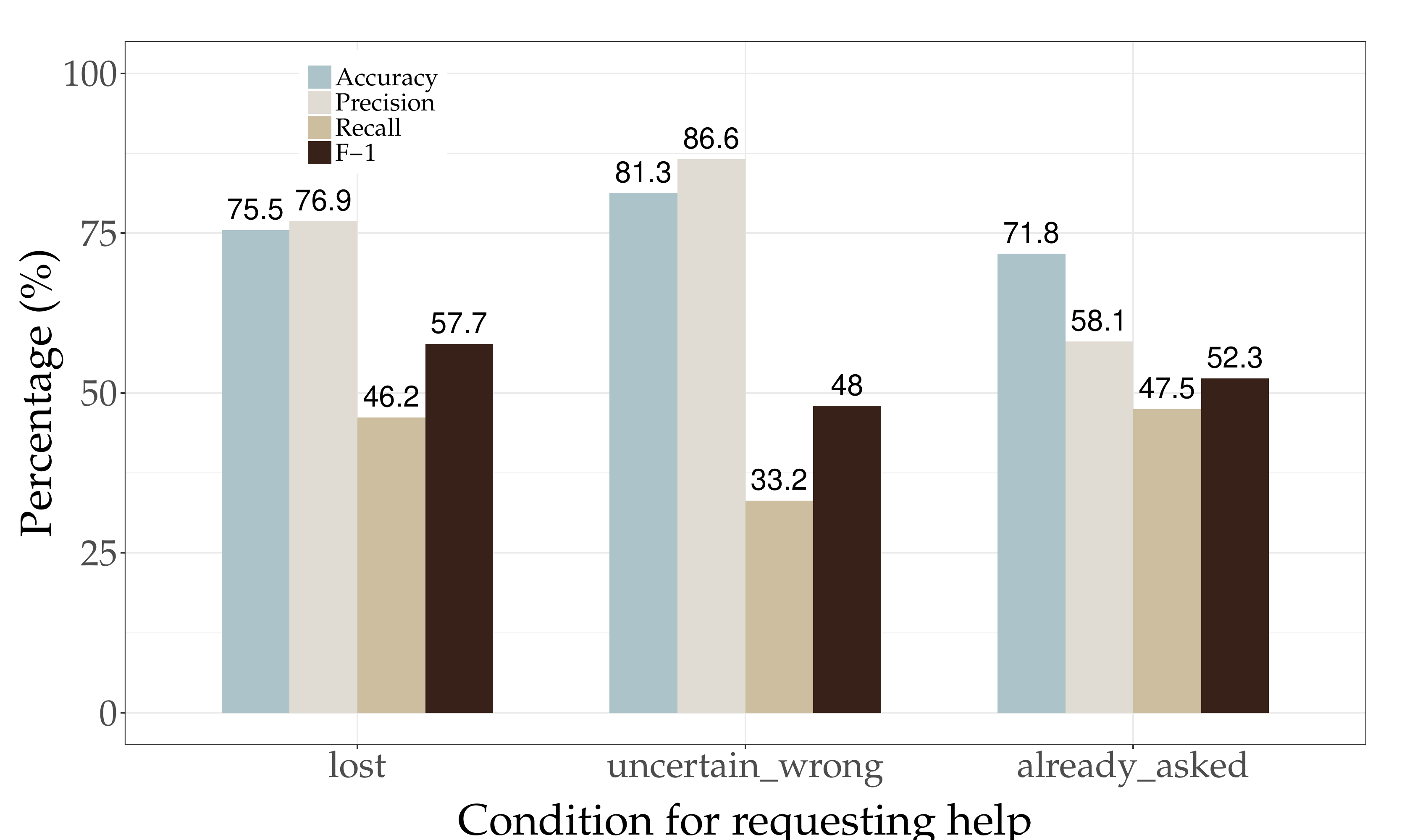}
    \caption{Accuracy, precision, recall, and F-1 scores in predicting the help-request conditions on \testSet \unseenAll. The \texttt{already\_asked} condition is the negation of the \texttt{never\_asked} condition.}
    \label{fig:ask_metrics}
\end{figure}

\noindent\textbf{Ablation Study}. \autoref{tab:ablation} shows an ablation study on the techniques we propose in this paper. We observe the performance on \valSet \unseenLang of the model trained without the curiosity-encouraging loss ($\alpha = 0$) is slightly higher than that of our final model, indicating that training with the curiosity-encouraging loss slightly hurts memorization.
This is expected because the model has relatively small size but has to devote part of its parameters to learn the curiosity-encouraging behavior.
On the other hand, optimizing with the curiosity-encouraging loss helps our agent generalize better to unseen environments. 
Predicting help-request conditions produces contrasting effects on the agent in seen and unseen environments, boosting performance in \valSet \unseenLang while slightly degrading performance in \valSet \unseenAll.
We investigate the help-request patterns and find that, in both types of environments, the agent makes significantly more requests when not learning to predict the conditions.
Specifically, the no-condition-prediction agent meets the \texttt{uncertain\_wrong} condition considerably more often (+5.2\% on \valSet \unseenLang and +6.4\% on \valSet \unseenAll), and also requests help at a previously visited location while executing the same task more frequently (+7.22\% on \valSet \unseenLang and +8.59\% on \valSet \unseenAll).
This phenomenon highlights a challenge in training the navigation and training the help-request policies jointly: as the navigation policy gets better, there are less positive examples (i.e., situations where the agent needs help) for the help-request policy to learn from. 
In this specific case, learning a navigation policy that is less accurate in seen environments is beneficial to the agent when it encounters unseen environments because such a navigation policy creates more positive examples for training the help-request.
Devising learning strategies that learns both efficient navigation and help-request policies is an exciting future direction. 

\noindent \textbf{Help-request Behavior on \testSet \unseenAll}. 
Our final agent requests about 18\% of the total number of time steps (\autoref{fig:ask_dist}). 
Overall, it learns a conservative help-request policy (\autoref{fig:reason_dist}).
\autoref{fig:ask_metrics} shows how accurate our agent predicts the help-request conditions. 
The agent achieves high precision scores in predicting the \texttt{lost} and \texttt{uncertain\_wrong} conditions (76.9\% and 86.6\%), but achieves lower recalls in all conditions (less than 50\%).
Surprisingly, it predicts the \texttt{already\_asked} condition less accurate than the other two, even though this condition is intuitively fairly simple to realize.

\begin{table*}[t!]
    \small
    \centering
    \setlength{\tabcolsep}{3pt}
	\begin{tabular}{l@{~~~~~~~~~}ccccp{0.5cm}cccc}
		\toprule[1pt]
	    & \multicolumn{4}{c}{\valSet \unseenLang} && \multicolumn{4}{c}{\valSet \unseenAll} \\
	    \cmidrule[.002em](lr{0.4em}){2-5}  \cmidrule[.002em](lr{0.4em}){7-10}
	    \multicolumn{1}{l}{Agent} & SR $\uparrow$ & SPL $\uparrow$ & Nav. $\downarrow$ & Requests/&
	    & SR $\uparrow$ & SPL $\uparrow$ & Nav.  $\downarrow$ & Requests/  \\
	    & (\%) & (\%) & Err. (m) & task $\downarrow$&
	    & (\%) & (\%) & Err. (m) & task $\downarrow$\\
	    \specialrule{.002em}{0.3em}{-0.8em} \\
	    %\addlinespace[0.4em]
	    Final agent & ~~86.06 & ~~62.49 & 1.48 & 2.7 && ~~45.39 & ~~23.33 & ~~8.21 & \textbf{4.5} \\ 
	    No inter-task reset & ~~83.62 & ~~60.44 & 1.44 & 3.2 && ~~40.60 & ~~19.89 & ~~8.26 & 7.3 \\ 
	    No condition prediction & ~~69.51 & ~~44.87 & 3.11 & 4.7 && ~~\textbf{45.72} & ~~\textbf{23.35} & ~~\textbf{6.56} & 8.1 \\ 
	    No cosine-similarity attention ($\beta = \vec 0$) & ~~80.44 & ~~56.63 & 1.89 & 3.7 && ~~38.69 & ~~19.86 & ~~8.81 & 8.5 \\
	    No curiosity-encouraging loss ($\alpha = 0$) & ~~\textbf{86.66} & ~~\textbf{69.66} & \textbf{1.28} & \textbf{2.0} && ~~39.92 & ~~21.33 & ~~8.66 & 7.7 \\
        \bottomrule[1pt] 
	\end{tabular}
	\caption{Ablation study on our proposed techniques. Models are evaluated on the validation splits and with a {\color{myred} batch size of 1}. Best results in each column are in bold.}
	\label{tab:ablation_bs1}
\end{table*}
\section{Mini-batch Resetting}
\label{sec:batch_reset}

\begin{table*}[t!]
    \small
    \centering
    \setlength{\tabcolsep}{3pt}
	\begin{tabular}{l@{~~~~~~~~~}ccccp{0.5cm}cccc}
		\toprule[1pt]
	    & \multicolumn{4}{c}{\unseenLang} && \multicolumn{4}{c}{\unseenAll} \\
	    \cmidrule[.002em](lr{0.4em}){2-5}  \cmidrule[.002em](lr{0.4em}){7-10}
	    \multicolumn{1}{l}{Agent} & SR $\uparrow$ & SPL $\uparrow$ & Nav. $\downarrow$ & Requests/&
	    & SR $\uparrow$ & SPL $\uparrow$ & Nav.  $\downarrow$ & Requests/  \\
	    & (\%) & (\%) & Err. (m) & task $\downarrow$&
	    & (\%) & (\%) & Err. (m) & task $\downarrow$\\
	    \specialrule{.002em}{0.3em}{-0.8em} \\
	    \textbf{Non-learning agents} \\
	    %\addlinespace[0.4em]
	            \randomWalk & ~~~~0.54 & ~~~~0.33 & 15.38 & 0.0 && ~~~~0.46 & ~~~~0.23 & ~~15.34 & 0.0 \\ 
	               \forward & ~~~~5.98 & ~~~~4.19 & 14.61 & 0.0 && ~~~~6.36 & ~~~~4.78 & ~~13.81 & 0.0 \\ 
	    \specialrule{.002em}{0.3em}{-0.8em} \\
	    \textbf{Learning agents} \\
	    %\addlinespace[0.4em]
	    No assistance & ~~17.21 & ~~13.76 & 11.48 & 0.0 && ~~~~8.10 & ~~~~4.23 & 13.22 & 0.0 \\ 
	    Learn to interpret assistance (ours) & ~~\textbf{86.67} & ~~\textbf{63.27} & ~~\textbf{1.63} & \textbf{2.7} && ~~\textbf{45.43} & ~~\textbf{25.02} & ~~\textbf{8.04} & \textbf{4.4} \\ 
	    \specialrule{.002em}{0.3em}{-0.8em} \\
	    \textbf{Skylines} \\
	    %\addlinespace[0.4em]
	    \shortest & 100.00 & 100.00 & ~~0.00 & 0.0 && 100.00 & 100.00 & ~~0.00 & 0.0\\ 
	    Perfect assistance interpretation & ~~90.19 & ~~68.76 & ~~1.16 & 2.4 && ~~79.01 & ~~54.86 & ~~2.65 & 3.0 \\
        \bottomrule[1pt] 
	\end{tabular}
	\caption{Results on test splits. The agent with ``perfect assistance interpretation" uses the teacher navigation policy ($\tnavpol$) to make decisions when executing a subtask from \system. Results of our final system are in bold. Models are evaluated with a {\color{myred}batch size of 1}.}
	\label{tab:main_bs1}
\end{table*}

\begin{table}[t!]
    \small
    \centering
	\begin{tabular}{lcc}
		\toprule[1pt]
		 \multicolumn{1}{l}{Assistance type}& \unseenLang & \unseenAll \\
		 \cmidrule(lr){1-1} \cmidrule(lr){2-2} \cmidrule(lr){3-3}
		Target image only & 84.29 & 29.91 \\
		+ Language instruction & \textbf{86.67} & \textbf{45.43}\\
        \bottomrule[1pt] 
	\end{tabular}
	\caption{Success rates (\%) of agents on test splits with different types of assistance. Models are evaluated with a {\color{myred}batch size of 1}.}
	\label{tab:assistance_type_bs1}
\end{table}

\begin{table}[t!]
    \small
    \centering
    \setlength{\tabcolsep}{4pt}
	\begin{tabular}{lccp{0cm}cc}
		\toprule[1pt]
		 & 
		 \multicolumn{2}{c}{\unseenLang} && 
		 \multicolumn{2}{c}{\unseenAll} \\
		 \cmidrule[.002em](lr){2-3}  \cmidrule[.002em](lr){5-6}
		 \multicolumn{1}{l}{$\aaskpol$} & SR $\uparrow$ & Requests/ && SR $\uparrow$ & Requests/ \\
		 & (\%) & task $\downarrow$ && (\%) & task $\downarrow$ \\
		 \cmidrule[.002em](lr){1-1}  \cmidrule[.002em](lr){2-3}  \cmidrule[.002em](lr){5-6}
		\noAsk & 17.21 & 0.0 && ~~8.10 & 0.0 \\
		\randomAsk & 78.40 & 4.1 && 35.97 & 6.6 \\
		\textsc{AskEvery5} & 86.63 & 3.5 && 33.62 & 7.1 \\
		Learned (ours) & \textbf{86.67} & \textbf{2.7} && \textbf{45.43} & \textbf{4.4} \\
        \bottomrule[1pt] 
	\end{tabular}
	\caption{Success rates (\%) of different help-request policies on test splits. Models are evaluated with a {\color{myred}batch size of 1}.}
	\label{tab:compare_askpol_bs1}
\end{table}

\begin{table}[t!]
    \small
    \centering
    \setlength{\tabcolsep}{3.1pt}
	\begin{tabular}{lccc}
		\toprule[1pt]
		 & SR $\uparrow$ & Nav. mistake $\downarrow$ & Help-request $\downarrow$ \\
		 Model & (\%) & repeat (\%) & repeat (\%)
		 \\ \midrule[0.002em] 
		\textsc{LSTM-EncDec} & 19.25 & 31.09 & 49.37 \\ 
		Our model ($\alpha = 0$) & 41.92 & 25.20 & 39.85 \\
		%Our model ($\alpha = 0$) & 39.29 & 20.75 & 40.77 \\
		Our model ($\alpha = 1$) & \textbf{45.43}& \textbf{20.53} & \textbf{10.80} \\ 
		%Our model ($\alpha = 1$) & \textbf{45.64}& \textbf{19.24} & \textbf{21.14} \\ 
        \bottomrule[1pt] 
	\end{tabular}
	\caption{Results on \testSet \unseenAll of our model, trained with and without curiosity-encouraging loss, and an LSTM-based encoder-decoder model (both models have about 15M parameters). ``Navigation mistake repeat" is the fraction of time steps on which the agent repeats a non-optimal navigation action at a previously visited location while executing the same task. 
	``Help-request repeat" is the fraction of help requests made at a previously visited location while executing the same task. Models are evaluated with a {\color{myred}batch size of 1}.}
	\label{tab:proposed_method_bs1}
\end{table}

Our implementation employs a \emph{mini-batch resetting} mechanism: every time an agent's inter-task module and local time need to be reset, we also reset the inter-task modules and local times of \emph{all} agents in the same mini-batch.
Mini-batch resetting provides the model another mechanism to escape looping situations by forcing it to forget the past, effectively boosting success rates by about 1-2\%.
However, it causes the agent to act non-derministically on a task if the batch size is greater than 1, because its behavior depends on behaviors of other agents that are grouped with it in the same mini-batch.
As a result, the success rate becomes dependent on the mini-batch construction.
This complicates comparison and evaluation of our methods. 

To simplify comparison and evaluation of our methods, in this section, we provide results obtain \emph{without} mini-batch resetting.
To obtain these results, we perform evaluation with a batch size of 1, hence eliminating the dependency on the mini-batch construction.
The results are slightly lower than those in \autoref{sec:results} and \autoref{sec:analysis} but all conclusions and qualitative claims made on those results remain true on both sets of results.
{\color{myred}\textbf{We strongly recommend future work reference results in this section when comparing with our methods}} (Tables \ref{tab:ablation_bs1}, \ref{tab:main_bs1}, \ref{tab:assistance_type_bs1}, \ref{tab:compare_askpol_bs1}, and \ref{tab:proposed_method_bs1}).

\end{document}